\documentclass[12pt]{article}
\usepackage{epsf}
\usepackage{bm,physics}
\usepackage{cite}
\usepackage{color}
\usepackage{amsmath,amssymb}
\usepackage{graphicx}
\usepackage[colorlinks,citecolor=blue]{hyperref}
\usepackage{comment}
\usepackage {diagbox} 
\renewcommand{\thefootnote}{\fnsymbol{footnote}}
\def\thefootnote{\fnsymbol{footnote}}

\makeatletter

\@addtoreset{equation}{section}
\makeatother

\begin{document}

\begin{titlepage}

\begin{center}

\vskip .45in

\bigskip\bigskip

{\Large \bf Exploring the Primordial Power Spectrum \\[8pt] with  Dark-Age 21\,cm fluctuations} 

\vskip .65in

{\large 
Fumiya~Okamatsu$^{1}$,
Tomo~Takahashi$^{2}$ and
Shuichiro Yokoyama$^{3,4,5}$
\vspace{2mm} \\
}
\vskip 0.2in

{\em 
$^{1}$Department of Physics, College of Humanities and Sciences, Nihon University, Tokyo 156-8550, Japan
\vspace{2mm}\\
$^{2}$Department of Physics, Saga University, Saga 840-8502, Japan
\vspace{2mm}\\
$^{3}$Kobayashi Maskawa Institute, Nagoya University, Aichi 464-8602, Japan
\vspace{2mm}\\
$^{4}$Department of Physics, Nagoya University, Nagoya, Aichi 464-8602, Japan
\vspace{2mm}\\
$^{5}$Kavli Institute for the Physics and Mathematics of the Universe (WPI), The University of Tokyo, Kashiwa, Chiba 277-8583, Japan
}

\end{center}
\vskip .5in

\begin{abstract}
We investigate the potential of future observations of the Dark Ages 21\,cm power spectrum to probe the scale dependence of the primordial power spectrum and distinguish between inflationary models. We find that the Dark Ages 21\,cm power spectrum can serve as a powerful probe of inflationary physics on scales beyond the reach of current CMB observations. While models predicting nearly identical running spectral indices may remain difficult to distinguish even with future 21\,cm observations, models with unconventional scale dependence can be stringently tested. We further show that combining 21\,cm observations with CMB data can significantly improve the constraints, even for a moderate observational configuration. We examine the impact of the choice of pivot scale on the resulting constraints.

\end{abstract}

\end{titlepage}

\renewcommand{\thepage}{\arabic{page}}
\setcounter{page}{1}
\renewcommand{\thefootnote}{\#\arabic{footnote}}
\setcounter{footnote}{0}

%%%%%%%%%%%%%%%%%%%%%%%%%%%%%%%%%%%%%%%%
\section{Introduction} \label{sec:intro}
%%%%%%%%%%%%%%%%%%%%%%%%%%%%%%%%%%%%%%%%

With the advent of high-precision cosmological observations, models of inflation have been tightly constrained, particularly through measurements of the spectral index $n_s$, characterizing the scale dependence of primordial fluctuations, and the tensor-to-scalar ratio $r$, quantifying the amplitude of primordial gravitational waves, which have been obtained from observations of cosmic microwave background (CMB) in combination with other observations such as baryon acoustic oscillation (BAO) \cite{Planck:2018jri,BICEP:2021xfz,AtacamaCosmologyTelescope:2025nti,SPT-3G:2025bzu}.
However, a definitive model of inflation has yet to be identified since multiple models can predict values of $n_s$ and $r$ that are consistent with current data. To break the degeneracy among competing models, additional observables beyond $n_s$ and $r$ are indispensable.

Among potential observables, the running of the spectral index and even its higher-order running have been proposed to capture the detailed scale-dependence of the spectral index $n_s$. To probe such running parameters, cosmological observations over a wide range must be examined. 
Although CMB observations place tight constraints on the spectral index, they are limited to large scales and thus measure the running parameters only mildly. To achieve more precise measurements of the detailed scale dependence, it is imperative to incorporate observations of smaller scales. Indeed several observations are put forward to probe small scales such as galaxy UV luminosity function \cite{Yoshiura:2020soa}, reionization  history \cite{Minoda:2023oqs}, gravitational lensing \cite{Gilman:2021gkj}, supernova lensing \cite{Ben-Dayan:2015zha},  21\,cm line of hydrogen from the cosmic dawn using the global signal \cite{Yoshiura:2018zts,  Yoshiura:2019zxq}, fluctuations from minihalos \cite{Sekiguchi:2017cdy} and intergalactic medium \cite{Mao:2008ug, Kohri:2013mxa}, 21\,cm forest \cite{Shimabukuro:2014ava}, Lyman-$\alpha$ forest \cite{Bird:2010mp, Palanque-Delabrouille:2015pga}, spectral distortions of CMB \cite{Khatri:2013dha, Clesse:2014pna, Cabass:2016ldu, Kainulainen:2017gqq}, substructure of dark matter halo \cite{Ando:2022tpj, Ando:2026xcs}, abundance of early galaxies \cite{Kobayashi:2026sji}, 
among others. 

One of the limitations of small-scale observations is the uncertainty introduced by astrophysics, which can degrade the precision of the constraints. However, the 21\,cm signal from the Dark Ages presents an exception, as this era is prior to the formation of the first astrophysical objects, which allows us to predict the signal without such an ambiguity. 
In fact, the 21\,cm signals from the Dark Ages fall into the frequency range that cannot be observed from ground telescopes due to the ionosphere and radio frequency interference (RFI). However,  several projects have been proposed with the aim of detecting the 21\,cm signal from the Dark Ages, either from the Moon or satellite orbiting it, including FARSIDE \cite{FARSIDEweb, Burns:2021pkx},  DAPPER \cite{DAPPERweb, Burns:2021ndk}, NCLE \cite{NCLEweb,2020AAS...23610203C}, LCRT \cite{LCRTweb, Goel:2022jgw}, DSL \cite{Chen:2019xvd}, TREED \cite{TREED}, TSUKUYOMI \cite{tsukuyomi}. Since the 21\,cm signal from the Dark Ages probes smaller scales, it provides an exceptional opportunity to investigate primordial fluctuations.  Some works such as Refs.~\cite{deKruijf:2024voc,Wen:2026kvc}
have explored how precise measurement of the 21\,cm power spectrum from this era can probe the primordial power spectrum. In this paper, we investigate the extent to which future observations of the 21\,cm power spectrum from the Dark Ages can constrain the running parameters and assess their potential to provide new insights into inflationary models\footnote{For studies probing other cosmological aspects, such as dark matter, with the Dark Ages 21\,cm power spectrum, see, e.g., \cite{Yamauchi:2022fri,Mondal:2023bxb,Park:2025phj,Sikder:2025ykc}. Furthermore, the global (spatially averaged) Dark Ages 21\,cm signal can also provide a valuable probe of cosmological parameters \cite{Mondal:2023xjx},  scenarios beyond the standard model \cite{Okamatsu:2023diy}, among other applications. For additional relevant studies, for example, see~\cite{Liu:2022iyy} and the references therein.
}. 

The structure of this paper is as follows. In Section~\ref{sec:21cmPS}, we give the formalism to compute 21\,cm power spectrum from the Dark Ages. In Section~\ref{sec:inflation}, we briefly review how we parametrize the primordial power spectrum with the running parameters. Several
examples of explicit inflation models are also discussed. 
In Section~\ref{sec:results}, first,
the methodology to investigate expected constraints from future 21\,cm power spectrum from the Dark Ages, in combination with the CMB, is described. Then,
we investigate the expected constraints on the spectral index $n_s$ and its running parameters $\alpha_s$ and $\beta_s$ from the Dark Ages 21\,cm power spectrum for several specifications of future experiments.
We also examine the impact of the choice of pivot scale on the resulting constraints.
We conclude this paper in the final section.

%%%%%%%%%%%%%%%%%%%%
\section{21\,cm power spectrum \label{sec:21cmPS}}
%%%%%%%%%%%%%%%%%%%%

In this section, we outline the formalism to compute the power spectrum of the 21\,cm fluctuation and the noise power spectrum, which will serve as the basis for studying expected constraints on the scale dependence of primordial power spectrum in the following sections. For the detailed description of the 21\,cm line, we refer the readers to \cite{Furlanetto:2006jb, Pritchard:2011xb}.

\subsection{21\,cm line signal}
The 21\,cm line of neutral hydrogen is a radio signal at a wavelength of 21\,cm or a frequency of 1420~MHz.
Here we give some basic formulas to calculate the 21\,cm signal, particularly during the Dark Ages. The 21\,cm signal is characterized by the differential brightness temperature~$T_b$: 
\begin{equation}
    \label{eq:Tb}
    T_{b}=\frac{T_{s}-T_{\gamma}}{1+z}\left(1-e^{-\tau_{\nu}}\right),
\end{equation}
where $T_s$ and $T_\gamma$ are the spin and the radiation temperatures, respectively. $\tau_{\nu}$ is the optical depth, which is written as 
\begin{equation}
    \label{eq:taunu}
\tau_{\nu}=\frac{3ch_{p}\lambda_{21}^{2}A_{10}x_{\rm HI}(1-Y_{p})n_{b}}{32\pi k_{B}T_{s}H}.
\end{equation}
Here, $\lambda_{21}$ is the wavelength of the 21\,cm line, $A_{10}$ is the Einstein coefficient for spontaneous emission, $x_{\rm HI}$ is the neutral hydrogen fraction, $Y_{p}$ is the primordial mass fraction of helium, $n_{b}$ is the number density of baryons, $H$ is the Hubble parameter, $c$ is the speed of light, $h_{p}$ is Planck's constant, and $k_{B}$ is the Boltzmann constant. 
For the case of interest, we can assume that $\tau_{\nu}\ll1$, and hence we can approximate Eq.~\eqref{eq:Tb} as
\begin{align}
    \label{eq:expand_Tb}
    T_b ~\simeq & ~\frac{T_s-T_{\gamma}}{1+z}\frac{3ch_{p}\lambda_{21}^{2}A_{10}x_{\rm HI}(1-Y_{p})n_{b}}{32\pi k_{B}T_{s}H}  \\
    \label{eq:Tb_cosmolopara}
    \simeq& ~85~{\rm mK}\left(\frac{T_s-T_{\gamma}}{T_s}\right)\left(\frac{\Omega_{b}h^2}{0.02237}\right) 
    \left(\frac{0.144}{\Omega_{m}h^2}\right)^{1/2}\left(\frac{1-Y_p}{1-0.24}\right)\left(\frac{1+z}{100}\right)
^{1/2}x_{\rm HI},
\end{align}
where $\Omega_{b}$ and $\Omega_{m}$ are the density parameters for baryons and matter, respectively. $h$ is the dimensionless Hubble constant defined via $H_0 = 100\,h~{\rm km}/ {\rm s} / {\rm Mpc}$.
The evolution of the spin temperature can be described by
\begin{equation}
    \label{eq:Ts}
    T_{s}^{-1}=\frac{T_{\gamma}^{-1}+x_{c}T_{K}^{-1}+x_{\alpha}T_{K}^{-1}}{1+x_{c}+x_{\alpha}},
\end{equation}
where $T_{K}$ is the gas temperature, $x_{c}$ and $x_{\alpha}$ are the coefficients for collisional couplings and the Wouthuysen-Field (WF) effect. Since Lyman-$\alpha$ photons are absent during the Dark Ages in the standard scenario, we neglect the WF effect in the following analysis. In this case, the evolution of the spin temperature~$T_s$ in the Dark Ages is characterized by the collisional coupling coefficient~$x_c$, which is determined by Hydrogen–Hydrogen (HH), electron-Hydrogen (eH), and proton-Hydrogen (pH) collisions (see, e.g., \cite{Pritchard:2011xb}):
\begin{align}
x_c
&= x_c^{\rm HH} + x_c^{\rm eH} + x_c^{\rm pH} \notag \\
& = \frac{T_{21}}{A_{10}T_\gamma} 
\qty[ \kappa^{\rm HH}(T_K)n_H +\kappa^{\rm eH}(T_K) n_e + \kappa^{\rm pH}(T_K) n_p] \,,
\end{align}
where $\kappa^{\rm HH}, \kappa^{\rm eH}$ and $\kappa^{\rm pH}$ are the scattering rates for HH, eH and pH collisions, respectively. 
$n_H, n_e$ and $n_p$ are the number densities of Hydrogen, electron and proton. In this work, we use the rates given in \cite{Furlanetto:2006jb, Furlanetto:2007te}.

After recombination, the residual electrons keep gas and radiation temperatures the same via Compton scattering. However, at redshift $z\sim 200$, the Compton scattering becomes ineffective at maintaining $T_K = T_\gamma$, and the gas temperature starts to drop faster than radiation, driving the spin temperature to $T_s\simeq T_K < T_\gamma$, so that $T_b < 0$ and the 21\,cm line appears in absorption. At later times, as the collisional coupling becomes inefficient due to cosmic expansion, the spin temperature gradually approaches the radiation temperature, i.e., $T_s \to T_\gamma$ and thus $T_b \to 0$. This process occurs during the Dark Ages, resulting in an absorption feature in the 21\,cm signal.

%%%%%%%%%%%%%%%%%
\subsection{Power spectrum of 21\,cm fluctuations}
%%%%%%%%%%%%%%%%%
Now we consider fluctuations of the 21\,cm line and describe how we can calculate its power spectrum.   We denote fluctuations of a quantity $X$ as  $\delta_X (z, \bm x) \equiv(X (z, \bm x)-\bar{X} (z))/\bar{X} (z)$, where $\bar{X} (z)$ represents its background component. 
Eq.~\eqref{eq:expand_Tb} indicates that fluctuations in the differential brightness temperature $T_b$ is determined by those of baryon density~$\delta_b$, ionization rate~$\delta_x$, Ly$\alpha$ coupling coefficient~$\delta_{\alpha}$, the line-of-sight peculiar velocity gradient~$\delta_v$, and the gas temperature~$\delta_T$. Thus we can express $\delta_{T_b}$, at a linear order \cite{Furlanetto:2006jb,Pritchard:2011xb}, as 
\begin{align}
    \label{eq:delta_Tb}
    \delta_{T_b}=\beta_{b}\delta_b + \beta_{x}\delta_x + \beta_{\alpha}\delta_{\alpha} + \beta_{T}\delta_{T}-\delta_v,
\end{align}
where $\beta_i$ is a coefficient that represents the degree to which each fluctuation contributes to $\delta_{T_b}$. Using the Fourier transform $\delta_{T_b}(z,  \bm{k})$ of the 21\,cm line fluctuation~$\delta_{T_b}(z,\bm{x})$, the power spectrum can be defined as follows:
\begin{align}
    \label{eq:PS}
    \langle \delta_{T_b}(z,\bm{k})\delta_{T_b}(z,\bm{k}~')\rangle=(2\pi)^3\delta_{D}^3(\bm{k}+\bm{k}^{'})P(z,\bm{k}).
\end{align}

To calculate the power spectrum, we define the monopole source as  \cite{Lewis:2007kz}
\begin{equation}
    \label{eq:monopole}
    \delta_{s}\equiv\delta_b+\frac{\bar{T}_{\gamma}}{\bar{T}_{s}-\bar{T}_{\gamma}}\left(\delta_{T_{s}}-\delta_{T_{\gamma }}\right) \,.
\end{equation}
Since the monopole and baryon density source contributions are dominant, the total spherically averaged 21\,cm power spectrum is expressed by
\begin{equation}
    \label{eq:total_PS}
    P(z, k)=\frac{1}{5}P_{\mu^4}(z, k)+\frac{1}{3}P_{\mu^2}(z, k)+P_{\mu^0}(z, k),
\end{equation}
where $P_{\mu^0}(z, k)$ is the monopole 21\,cm power spectrum, $P_{\mu^4}(z, k)$ is the power spectrum of the baryon density, and $P_{\mu^0}(z, k)$ is the cross term of the 21\,cm fluctuation and the baryon density fluctuation. We also define the dimensionless power spectrum:
\begin{equation}
    \label{eq:dimensionless_PS}
    \Delta_{21}\equiv\frac{k^3}{2\pi^2}P(z, k).
\end{equation}
Illustrative examples of the power spectra for $\nu = 20, 30$ and $40~{\rm MHz}$ are presented in Fig.~\ref{fig:PS_noise}, to be compared with the noise spectrum described in the next subsection.
For the cosmological parameters, we set the baryon and dark matter densities
$\Omega_{b}h^2=0.02237$, $\Omega_{c}h^2=0.12$, the Hubble constant $h=0.6736$, the amplitude and the spectral index of primordial power spectrum
$A_s=2.1\times10^{-9}$ and $n_s=0.9665$, the optical depth
$\tau = 0.0544$ which are the mean values from
Planck data \cite{Planck:2018vyg}, and $Y_{p}=0.2436$ from Hysu {\em et al} \cite{Hsyu:2020uqb}.

%%%%
\subsection{Noise power spectrum
\label{sec:noise_ps}}
%%%%

In the analysis, we consider two contributions to the noise power spectrum. The first one is the uncertainty due to cosmic variance (CV) \cite{Mondal:2015oga},
which is given by 
\begin{equation}
    \label{eq:CV_error}
    \delta P_{\rm CV}(\nu, k)=\frac{2\pi P(\nu, k)}{\sqrt{V(\nu)k^{3}\Delta(\ln k)}},
\end{equation}
where $k$ is the central wave number in the bin, 
$V(\nu)=\Omega_{\rm FoV}r_{\nu}^{2}\Delta r_{\Delta\nu}$ is the survey volume for the frequency (redshift) bin, $\Omega_{\rm FoV}=\frac{[21(1+z)~{\rm cm}]^{2}}{A_{\rm eff}}$ is the field of view, $A_{\rm eff}$ is the effective collecting area of an antenna, $r_{\nu}(z)$ is the comoving distance to the center of the frequency bin and $\Delta r_{\Delta\nu}$ is the comoving length corresponding to the bandwidth $\Delta\nu$. The second contribution is the thermal noise, which is dominant for the scales of our interest. 
The thermal noise is given by 
\cite{Mellema:2012ht}
\begin{equation}
    \label{eq:thermal_noise}
    \delta P_{\rm thermal}(\nu, k) = \frac{2}{\pi}\left(\frac{k^{3}V(\nu)}{\Delta(\ln k)}\right)^{1/2}\frac{T_{\rm sys}^{2}}{\Delta\nu t_{\rm int}}\frac{1}{N^{2}}\frac{A_{\rm core}}{A_{\rm eff}},
\end{equation}
where $N$ is the total number of stations and $A_{\rm core}=N\times A_{\rm eff}$ is the core area of the telescope array. The system temperature $T_{\rm sys}$ is approximately equal to the sky brightness temperature $T_{\rm sky} = 180\times(\nu/180~{\rm MHz})^{-2.6}~{\rm K}$ \cite{Furlanetto:2006jb}. 

\begin{figure}
    \centering
    \includegraphics[width=1.0\linewidth]{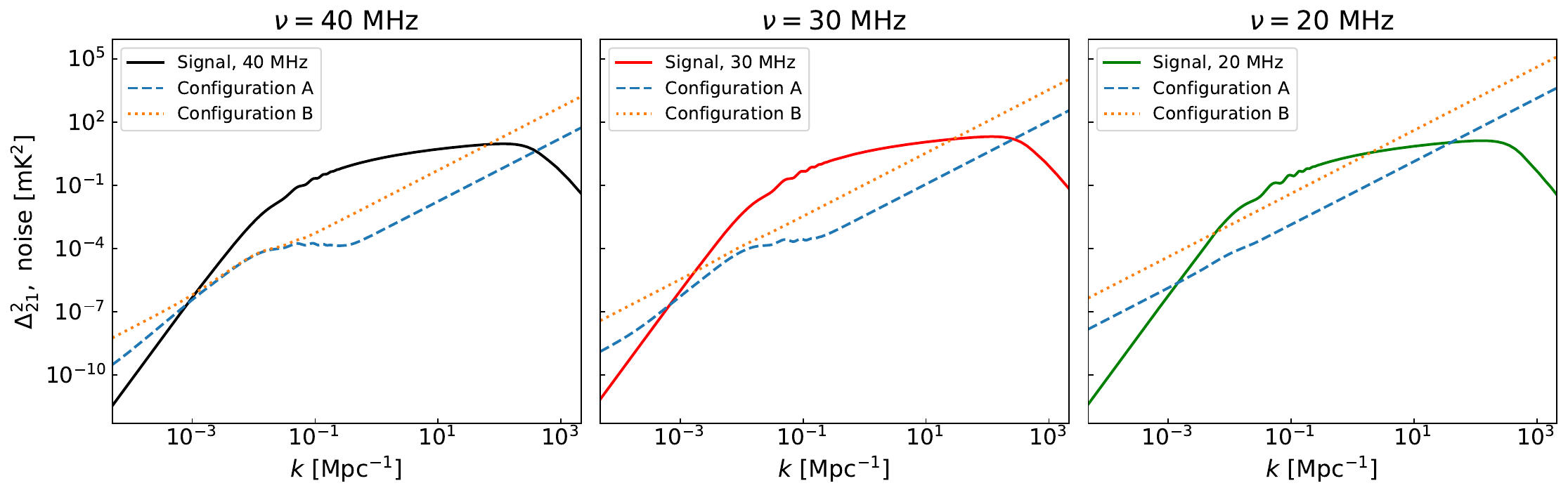}
    \caption{Plots of the 21\,cm power spectrum and the noise. Each solid line represents the 21\,cm power spectrum at $\nu=20,~30,~40$ MHz. In each dashed line, we show errors at $\nu = 20,~30,~40$ MHz for each observation configuration.}
    \label{fig:PS_noise}
\end{figure}

In the following analysis, we assume two configurations, denoted as configuration A and B,  as illustrative examples of future observations by specifying two parameters: the number of stations $N$ and the integration time $t_{\rm int}$. The values of these parameters for the configurations A and B are summarized in Table~\ref{tab:observation_spec}.
The number of stations is related to the effective area $A_{\rm eff}$ and the core area of the telescope array $A_{\rm core}$ via $A_{\rm core} = N\times A_{\rm eff}$. We set $A_{\rm eff}=25~{\rm m}^2$, and hence  the configurations A and B correspond to $A_{\rm core}=1000,~100~{\rm km}^2$, respectively.
Fig.~\ref{fig:PS_noise} compares the 21\,cm fluctuation and the noise power spectra at specific frequencies (redshifts) for each configuration. As seen from the figure, particularly for the frequency of $\nu=40~{\rm MHz}$, a broad range of scales can be probed as $6\times10^{-4}\le k\, [{\rm Mpc}^{-1}] \le 700$ for configuration A, which can give a severe constraint. Therefore, we adopt mock data consisting of 11 $k$ bins in the range $6\times10^{-4}\le k\, [{\rm Mpc}^{-1}] \le 700$, and 8 frequency bins in the range $10\le \nu\,[{\rm MHz}]\le 45$ ($30\lesssim\, z\lesssim\,140$) range.

\begin{table}
    \centering
    \begin{tabular}{|c|c|c|} \hline
         & Number of stations ($N$) & Integration time ($t_{\rm int}$)\\ \hline
         configuration A & $N=4\times10^7$ & $t_{\rm int}=3~{\rm years}$\\ \hline 
         configuration B& $N=4\times10^6$ & $t_{\rm int}=1~{\rm years}$ \\ \hline
    \end{tabular}
    \caption{
    Specifications of the observational configurations adopted in this paper.
    }
    \label{tab:observation_spec}
\end{table}

%%%%%%%%%%%%%%%%%%%%%%%%%%%%%%%%%%%%%%%%%
\section{Inflationary predictions \label{sec:inflation}}
%%%%%%%%%%%%%%%%%%%%%%%%%%%%%%%%%%%%%%%%%

In this section, we summarize the predictions of inflationary models, focusing on the scale dependence of the primordial power spectrum: the spectral index $n_s$, the running parameter $\alpha_s$, and the running of the running parameter $\beta_s$. First, we present a common parametrization of the primordial power spectrum in terms of these parameters, and then discuss several example models.

\subsection{\label{sec:general_formulas} General formulas}
The primordial power spectrum is commonly described by a power-law form as follows:
\begin{equation}
    \label{eq:PPS}
    \mathcal{P}_{s}(k)=A_s\left(\frac{k}{k_{\ast}}\right)^{n_s-1+\frac{1}{2}\alpha_s\ln(k/k_{\ast})+\frac{1}{3!}\beta_s\ln^2(k/k_{\ast})},
\end{equation}
where $A_s$  and $n_s$ are the amplitude and the spectral index at the pivot scale $k_{\ast}$, respectively.
In this parameterization, we have also incorporated the scale-dependence of $n_s$ in Eq.~\eqref{eq:PPS}, which can be described by the so-called the running and the running of the running parameters, which are defined as 
\begin{align}
\label{eq:running_def}
  \alpha_s\equiv \frac{dn_s}{d\ln k} \,, 
  \qquad\qquad
  \beta_s\equiv \frac{d\alpha_s}{d\ln k} \,.
\end{align}

Assuming a single-field slow-roll inflation model with a canonical kinetic term, $n_s$, $\alpha_s$, and $\beta_s$ can be written, by using the slow-roll parameters, as follows: 
\begin{align}
    \label{eq:ns_slow_para}
    n_s-1 &= -6\epsilon+2\eta,\\
    \label{eq:alpha_slow_para}
    \alpha_s &=-24\epsilon^2+16\epsilon\eta-2\xi^{(2)}, \\
    \label{eq:beta_slow_para}
    \beta_s &= -192\epsilon^3+192\epsilon^2\eta-32\epsilon\eta^2-24\epsilon\xi^{(2)}+2\eta\xi^{(2)}+2\sigma^{(3)},
\end{align}
where the slow-roll parameters are defined, with respect to an inflaton potential $V(\phi)$, as
\begin{align}
    \label{eq:slow_para_single}
    \epsilon&\equiv\frac{1}{2}M_{\rm pl}^2\left(\frac{V^{\prime}}{V}\right)^2,
    \qquad 
    \eta\equiv M_{\rm pl}^2\frac{V^{\prime\prime}}{V},\nonumber\\ \\
    \xi^{(2)}&\equiv M_{\rm pl}^4\frac{V^{\prime}V^{\prime\prime\prime}}{V^2},
    \qquad
    \sigma^{(3)}\equiv M_{\rm pl}^{6}\frac{(V^{\prime})^2V^{(4)}}{V^3}. \nonumber
\end{align}
Here, a prime represents derivatives with respect to $\phi$, and $V^{(4)}$ is the fourth-order derivative with respect to $\phi$. $M_{\rm pl}$ is the Planck mass. Given the potential for the inflaton, we can calculate the slow-roll parameters from Eq.~\eqref{eq:slow_para_single}, and then evaluate $n_s$, $\alpha_s$, and $\beta_s$ by using the above formulas. 

The formulas for $n_s, \alpha_s$ and $\beta_s$ given in Eqs.~\eqref{eq:ns_slow_para}--\eqref{eq:beta_slow_para} are only applicable to the case of single-field inflation models. Indeed, high-energy theories such as string theory may motivate us to consider multifield scenarios for the inflationary Universe. Here we consider a two-field case in which a spectator field $\chi$ is also present during inflation and contributes to primordial fluctuations. We assume that $\chi$ does not affect the background inflationary dynamics,  as in the curvaton scenario \cite{Enqvist:2001zp,Lyth:2001nq,Moroi:2001ct} and the modulated reheating model \cite{Dvali:2003em,Kofman:2003nx}, and that  the inflaton alone determines the background dynamics. Such a model is referred to as a mixed inflaton-spectator model and its predictions for the inflationary observables have been discussed in \cite{Langlois:2004nn,Moroi:2005kz,Moroi:2005np,Ichikawa:2008iq,Ichikawa:2008ne,Enqvist:2013paa,Morishita:2022bkr,Jinno:2023bpc}. The primordial power spectrum in the model is given by
\begin{align}
    \label{eq:multi_PPS}
    \mathcal{P}_s(k) = \mathcal{P}_s^{(\phi)}(k) + \mathcal{P}_s^{(\chi)}(k),
\end{align}
where 
$\mathcal{P}_s^{(\phi)}(k)$ and $ \mathcal{P}_s^{(\chi)}(k)$ 
represent the contributions from the inflaton $\phi$ and spectator $\chi$, respectively. 
We define the fractional contributions of the inflaton and spectator fields at the pivot scale $k_\ast$ as
\begin{align}
    Q_\phi = \frac{\mathcal{P}_s^{(\phi)}(k_\ast)}{\mathcal{P}_s^{(\phi)}(k_\ast) + \mathcal{P}_s^{(\chi)}(k_\ast)}\,,
    \qquad
    Q_\chi = \frac{\mathcal{P}_s^{(\chi)}(k_\ast)}{\mathcal{P}_s^{(\phi)}(k_\ast) + \mathcal{P}_s^{(\chi)}(k_\ast)}\,,
    \qquad 
    Q_{\phi} + Q_{\chi} = 1.
\end{align}
Since the inflaton and spectator contributions generally have different scale dependence, the effective spectral index and its runnings are determined from the total scalar power spectrum rather than from either component separately. 
Describing the scale dependence for $\mathcal{P}_s^{(\phi)}(k)$ and $ \mathcal{P}_s^{(\chi)}(k)$ respectively as
\begin{align}
    \mathcal{P}_s^{(\phi)}(k) & \propto \qty(\frac{k}{k_\ast})^{n_s^{(\phi)}-1+\frac{1}{2}\alpha_s^{(\phi)}\ln(k/k_{\ast})+\frac{1}{3!}\beta_s^{(\phi)}\ln^2(k/k_{\ast})} \,, \\[8pt]
    \mathcal{P}_s^{(\chi)}(k) & \propto \qty(\frac{k}{k_\ast})^{n_s^{(\chi)}-1+\frac{1}{2}\alpha_s^{(\chi)}\ln(k/k_{\ast})+\frac{1}{3!}\beta_s^{(\chi)}\ln^2(k/k_{\ast})} \,,
\end{align}
where $n_s^{(\phi)}, \alpha_s^{(\phi)}$ and $\beta_s^{(\phi)}$ are given by the same expressions as in Eqs.~\eqref{eq:ns_slow_para}, \eqref{eq:alpha_slow_para} and \eqref{eq:beta_slow_para}, respectively. On the other hand, the counterparts for the spectator contribution, $n_s^{(\chi)}, \alpha_s^{(\chi)}$ and $\beta_s^{(\chi)}$,  are written as follows:
\begin{align}
    n_s^{(\chi)} - 1 &= -2\epsilon + 2\eta_\chi, \label{eq:ns_chi}\\[8pt]
    \alpha_s^{(\chi)} &= -8\epsilon^2 + 4\epsilon\eta + 4\epsilon\eta_\chi - 2\xi_{\chi}^{(2)},\label{eq:alphas_chi}\\[8pt]
    \beta_s^{(\chi)} &= -64\epsilon^3 + 56\epsilon^2\eta + 24\epsilon^2\eta_\chi -8\epsilon\eta^2 -8\epsilon\eta\eta_\chi - 4\epsilon\xi^{(2)} - 12\epsilon\xi_\chi^{(2)} + 2\eta_\chi\xi_\chi^{(2)} + 2\sigma_\chi^{(3)}\label{eq:betas_chi} \,,
\end{align}
where $\eta_\chi, \xi^{(2)}_\chi$ and $\sigma^{(3)}_\chi$ are the slow-roll parameters for the spectator field, defined as
\begin{align}
    \label{eq:slowroll_spectator}
    \eta_\chi \equiv \frac{U''}{3H_\ast},\qquad\xi_{\chi}^{(2)}\equiv\frac{U'U'''}{(3H_\ast^2)^2},\qquad\sigma_{\chi}^{(3)} \equiv \frac{(U')^2U^{(4)}}{(3H_\ast^2)^3}
    \,.
\end{align}
Here $U$ is a potential for the spectator field and $H_\ast$ represents the Hubble parameter at the horizon exit during inflation.
In the subsequent analysis, we adopt the potential for the spectator field as
\begin{align}
\label{eq:U_spectator}
U(\chi) = \frac12 m_\chi^2\chi^2 \,,
\end{align}
where $m_\chi$ is the mass of the spectator field. With this potential, higher order slow-roll parameters for $\chi$ vanish: $\xi_\chi^{(2)} = \sigma_\chi^{(3)} = 0$.
We assume that $m_\chi$ is much smaller than $H_\ast$, which yields $\eta_\chi \simeq 0$.

The effective spectral index and its running are written as\footnote{
The spectral index for the total primordial power spectrum is defined as
\begin{align}
    n_s^{\rm (tot)}-1 = \frac{d \log \qty({\cal P}_s^{(\phi)}+ {\cal P}_s^{(\sigma)})}{d\log k} \,,
\end{align}
and the expressions for $\alpha_s$ and $\beta_s$ follow from Eq.~\eqref{eq:running_def}. Although the spectral index and its runnings defined above do not strictly coincide with $n_s^{\rm eff}$, $\alpha_s^{\rm eff}$, and $\beta_s^{\rm eff}$, the latter give the same values to a very good approximation when $n_s-1$, $\alpha_s$, and $\beta_s$ are sufficiently small, as is the case of interest in this paper.
} 
\begin{align}
    n_s^{\rm eff} - 1 &= Q_\phi (n_s^{(\phi)} - 1) + Q_\chi \left(n_s^{(\chi)} - 1\right) \label{eq:ns_eff},\\[8pt]
    \alpha_s^{\rm eff} &= Q_\phi \alpha_s^{(\phi)} + Q_\chi \alpha_s^{(\chi)} + Q_\phi Q_\chi \left(n_s^{(\phi)} - n_s^{(\chi)}\right)^2 \label{eq:alphas_eff},\\[8pt]
    \beta_s^{\rm eff} &= Q_\phi \beta_s^{(\phi)} + Q_\chi \beta_s^{(\chi)} + 3Q_\phi Q_\chi \left(n_s^{(\phi)} - n_s^{(\chi)}\right)\left(\alpha_s^{(\phi)} - \alpha_s^{(\chi)}\right) - Q_\phi Q_\chi (Q_\phi - Q_\chi) \left(n_s^{(\phi)} - n_s^{(\chi)}\right)^3 \label{eq:betas_eff} \,.
\end{align}
The tensor-to-scalar ratio is likewise modified because the scalar power spectrum receives contributions from both fields, which can be written as
\begin{align}
    r^{\rm eff} & = 16\epsilon Q_\phi.\label{eq:r_eff}
\end{align}
For mixed inflaton-spectator models considered in the following, we use these expressions to calculate the inflationary predictions.

\subsection{Example models}
\label{subsec:example_model}

Current precision measurements of cosmological observations, including the CMB in combination with BAO and Type Ia supernovae (SNeIa), place severe constraints on inflation models. Nevertheless, a variety of models remain consistent with current observations, and many of them exhibit a degeneracy in their predictions for $n_s$ and $r$. Even when model parameters are tuned to yield the same values of $n_s$ and $r$, their predictions for the running parameters, $\alpha_s$ and $\beta_s$, can differ. To illustrate this degeneracy and investigate the extent to which future dark-ages 21\,cm observations can distinguish between such models by using the runnings, 
we consider several representative inflation models and compare their predictions for the observables $n_s$, $r$, $\alpha_s$, and $\beta_s$.

\subsubsection{$\alpha$-attractor E-model}

Among the inflationary models consistent with current observational data, $\alpha$-attractor models are particularly well studied in recent years. Here, we consider the so called E-model \cite{Carrasco:2015rva} as an example of such an $\alpha$-attractor model, whose potential is given by
\begin{equation}
    V (\phi) = V_0 \left[ 1 - \exp \left( -\sqrt{\frac{2}{3\alpha}} \frac{\phi}{M_{\rm pl}} 
    \right)  \right]^{2m}
    = V_0 \left[ 1 - \exp \left( -\Phi
    \right)  \right]^{2m} \,,
\end{equation}
where $V_0$ represents the energy scale of the potential, and we have introduced a rescaled field value as $\Phi \equiv \sqrt{2/(3\alpha)}  (\phi /M_{\rm pl})$. The slow-roll parameters in this model are calculated as follows:
\begin{equation}
\label{eq:Emode_SR}
    \begin{aligned}
    & \epsilon = \frac{4m^2}{3\alpha ( e^\Phi -1)^2} \,, \qquad\qquad
    \eta = \frac{4 m \left(e^{\Phi }-2 m\right)}{3 \alpha  \left(e^{\Phi }-1\right)^2}
    \,,  \\
    & \xi^{(2)} = \frac{16 m^2 \left(4 m^2+(1-6 m) e^{\Phi }+e^{2 \Phi }\right)}{9 \alpha ^2 \left(e^{\Phi
   }-1\right)^4}    \,, \\[8pt]
    \sigma^{(3)} &= -\frac{64 m^3 \left(-8 m^3+\left(24 m^2-8 m+1\right) e^{\Phi }+(4-14 m) e^{2 \Phi }+e^{3
   \Phi }\right)}{27 \alpha ^3 \left(e^{\Phi }-1\right)^6}  \,.
\end{aligned}
\end{equation}
The field value at the end of inflation ($\epsilon_{\rm end}=1$) $\Phi_{\rm end}$ is given by 
\begin{align}
    e^{\Phi_{\rm end}}=1+\frac{2m}{\sqrt{3\alpha}}.
\end{align}
On the other hand, 
the value of the inflaton at the horizon exit $\phi_\ast$ can be expressed by the number of $e$-folds at the time when the pivot scales exited the horizon, denoted by $N_\ast$, using 
\begin{equation}
\label{eq:N_ast}
    N_\ast = \frac{1}{M_{\rm pl}^2} \int_{\phi_{\rm end}}^{\phi_\ast }
    \frac{V}{V'} \, d\phi \,,
\end{equation}
where $\phi_{\rm end}$ is the inflaton field value at the end of inflation.
Then, the slow-roll parameters can be written with respect to $N_\ast$.
From Eq.~\eqref{eq:N_ast}, in the $\alpha$-attractor E-model,
the relation between $N_\ast$ and $\Phi_\ast$ can be written as
\begin{align}
    N_\ast = \frac{3\alpha}{4m}\left[e^{\Phi_\ast}-\Phi_\ast-\left(1+\frac{2m}{\sqrt{3\alpha}}\right)+\ln\left(1+\frac{2m}{\sqrt{3\alpha}}\right)\right] \,.
\end{align} 
Although we use the exact relation between $\Phi_\ast$ and $N_\ast$ given above to predict inflationary observables such as $n_s$, $r$, $\alpha_s$, and $\beta_s$, a rough estimate can instead be obtained by expanding the relevant expressions in powers of $1/N_\ast$ as
\begin{align}
\label{eq:Emode_Nexpansion}
    n_s  \simeq 1 - \frac{2}{N_\ast} \,, 
    \qquad
    \alpha_s \simeq -\frac{2}{N_\ast^2} \,,
    \qquad
    \beta_s \simeq -\frac{4}{N_\ast^3} \,,
    \qquad
    r \simeq \frac{12 \alpha}{N_\ast^2} \,.
\end{align}

\subsubsection{Polynomial $\alpha$-attractor}

We also consider a yet another model for the $\alpha$-attractor type: the polynomial $\alpha$-attractor model \cite{Kallosh:2022feu}. The potential of this model is the following: 
\begin{equation}
    V (\phi) = \frac{ V_0 |\phi|^2}{\mu^2 + |\phi|^2} \,,
\end{equation}
where $V_0$ and $\mu$ are the model parameters. When we assume the scalar field is real, then $|\phi|^2=\phi^2$. The slow-roll parameters in this model are given by
\begin{equation}
    \begin{aligned}
    & \epsilon  =  \frac{2M_{\rm Pl}^2\mu^4}{\phi^2(\mu^2+\phi^2)^2}\,,
    \qquad
    \eta = \frac{2M_{\rm Pl}^2\mu^2(\mu^2-3\phi^2)}{\phi^2(\mu^2+\phi^2)^2} \,,
    \\[8pt]
    & \xi^{(2)} = \frac{48M_{\rm Pl}^4\mu^4(\phi^2-\mu^2)}{\phi^2(\mu^2+\phi^2)^4} \,,
    \qquad
     \sigma^{(3)} = -\frac{96M_{\rm Pl}^6\mu^6(\mu^4-10\mu^2\phi^2+5\phi^4)}{\phi^4(\mu^2+\phi^2)^6} \,.
\end{aligned}
\end{equation}
$\phi_{\rm end}$ can be evaluated by solving $\epsilon=1$, namely  
\begin{align}
    \phi_{\rm end}^2(\mu^2+\phi_{\rm end}^2)=2M_{\rm Pl}^2\mu^4.
\end{align}
The field value $\phi_\ast$ can be represented with $N_\ast$,
by using Eq.~\eqref{eq:N_ast},  as
\begin{align}
    \phi_\ast^2=-\mu^2+\sqrt{(\mu^2+\phi_{\rm end}^2)^2+8M_{\rm Pl}^2\mu^2N_\ast}.
\end{align}
Although we use this relation in our calculations to predict the inflationary observables, approximate expressions for $n_s,~r,~\alpha_s$, and $\beta_s$ in the limit of a large $N_\ast$, analogous to Eq.~\eqref{eq:Emode_Nexpansion} are given as:
\begin{equation}
    n_s  \simeq 1 - \frac{3}{2N_\ast} \,, 
    \qquad
    \alpha_s \simeq -\frac{3}{2N_\ast^2} \,,
    \qquad
    \beta_s \simeq -\frac{3}{N_\ast^3} \,,
    \qquad
    r \simeq \frac{\sqrt{2}\mu}{M_{\rm Pl}}\frac{1}{N_\ast^{3/2}} \,.
\end{equation}

\subsubsection{Hilltop model}
Another single-field model which is consistent with current observations is Hilltop inflation model \cite{Boubekeur:2005zm}.
The potential of this model is given by
\begin{equation}
    V (\phi) = V_0 \left[ 1 - \left( \frac{\phi}{\mu}\right)^p  \right] \,,
\end{equation}
where $\mu$ and $p$ are the model parameters. The slow roll parameters in this model are calculated as
\begin{equation}
\label{eq:hilltop_SR}
    \begin{aligned}
    \epsilon  &= \frac{p M_{\rm Pl}^2}{2\mu^2}\frac{(\phi/\mu)^{2p-2}}{\left[1-(\phi/\mu)^p\right]^2} \,,
    \qquad
    \eta = -\frac{p(p-1)M_{\rm Pl}^2}{\mu^2}\frac{(\phi/\mu)^{p-2}}{\left[1-(\phi/\mu)^p\right]} \,,
    \\[8pt]
    \xi^{(2)} &= \frac{p^2(p-1)(p-2)M_{\rm Pl}^4}{\mu^4}\frac{(\phi/\mu)^{2p-4}}{\left[1-(\phi/\mu)^p\right]^2} \,,
    \qquad
    \sigma^{(3)} = -\frac{p^3(p-1)(p-2)(p-3)M_{\rm Pl}^6}{\mu^6}\frac{(\phi/\mu)^{3p-6}}{\left[1-(\phi/\mu)^p\right]^3} \,. \\[8pt]
\end{aligned}
\end{equation}
The inflaton value at the end of inflation $\phi_{\rm end}$ can be obtained by solving the following equation:
\begin{align}
    \left(\frac{\phi_{\rm end}}{\mu}\right)^{p-1}=\frac{\sqrt{2}\mu}{pM_{\rm Pl}}\left[1-\left(\frac{\phi_{\rm end}}{\mu}\right)^p\right],
\end{align}
and it can be approximately written in a simple form, for $p>2$ and $\phi_{\rm end} \ll \mu$, as
\begin{align}
    \phi_{\rm end}\simeq\mu\left[\frac{\mu^2}{p(p-1)M_{\rm Pl}^2}\right]^{1/(p-2)}.
\end{align}
The number of $e$-folds, $N_\ast$, is expressed as a function of the inflaton field value at the corresponding epoch, $\phi_\ast$, as follows:
\begin{align}
    N_\ast =
    \begin{cases}
        &\displaystyle\frac{\mu^2}{pM_{\rm Pl}^2}\left[\frac{(\phi_\ast/\mu)^{2-p}-(\phi_{\rm end}/\mu)^{2-p}}{p-2}-\frac{1}{2}\left\{\left(\frac{\phi_{\rm end}}{\mu}\right)^2-\left(\frac{\phi_{\ast}}{\mu}\right)^2\right\}\right] \,\,\, (p\neq2) \\
        \\
        &\displaystyle\frac{\mu^2}{pM_{\rm Pl}^2}\left[\ln\left(\frac{\phi_{\rm end}}{\phi_\ast}\right)-\frac{1}{2}\left\{\left(\frac{\phi_{\rm end}}{\mu}\right)^2-\left(\frac{\phi_{\ast}}{\mu}\right)^2\right\}\right].~~~~~~~~~~~~~~(p=2)
    \end{cases}
\end{align}
In the limit of a large $N_\ast$ and $p>2$, $n_s,~r,~\alpha_s$, and $\beta_s$ can be written as follows:
\begin{equation}
\begin{aligned}
    & n_s  \simeq 1 - \frac{2(p-1)}{(p-2)N_\ast} \,, 
    \quad
    \alpha_s \simeq -\frac{2(p-1)}{(p-2)N_\ast^2} \,,
    \quad
    \beta_s \simeq -\frac{4(p-1)}{(p-2)N_\ast^3} \,,
    \\[8pt]
     & r \simeq \frac{8p^2M_{\rm Pl}^2}{\mu^2}\left[\frac{\mu^2}{p(p-2)M_{\rm Pl}^2N_\ast}\right]^{\frac{2p-2}{p-1}} \,.
\end{aligned}
\end{equation}

\subsubsection{Natural-Spectator model}
In addition to single-field models, we also consider multi-field models, specifically, mixed inflaton-spectator models where both the inflaton and the spectator field can contribute to primordial fluctuations. The scale-dependence of the primordial power spectrum in this setup can be characterized by $n_s^{\rm eff}, \alpha_s^{\rm eff}$ and $\beta_s^{\rm eff}$ given in Eqs.~\eqref{eq:ns_eff}--\eqref{eq:betas_eff}, and the tensor-to-scalar ratio $r^{\rm eff}$ is given by Eq.~\eqref{eq:r_eff}.

As mentioned in Section~\ref{sec:general_formulas}, the potential for the spectator field $\chi$ is assumed to be given by Eq.~\eqref{eq:U_spectator}, with its mass much smaller than the Hubble rate during inflation, so that the slow-roll parameter for $\chi$ is negligible. For the inflaton potential, we assume that for natural inflation \cite{Freese:1990rb,Adams:1992bn}:
\begin{align}
    \label{eq:NS_V}
    V(\phi) = \Lambda^4\left[1 - \cos\left(\frac{\phi}{f}\right)\right],\qquad 0\le\phi\le\pi f,
\end{align}
where $\Lambda$ and $f$ denote the energy scale of the model and some breaking scale which determines the curvature of the potential, respectively. The slow-roll parameters in this model are calculated as follows:
\begin{equation}
    \label{eq:slow_roll_NS}
        \begin{aligned}
    \epsilon &= \frac{M_{\rm Pl}^2}{2f^2}\frac{\cos^2(\phi_\ast/2f)}{\sin^2(\phi_\ast/2f)}, \qquad
    \eta = \frac{M_{\rm Pl}^2}{f^2}\frac{1 - 2\sin^2(\phi_\ast/2f)}{2\sin^2(\phi_\ast/2f)},\\
    \xi^{(2)} &= -\frac{M_{\rm Pl}^4}{f^4}\frac{\cos^2(\phi_\ast/2f)}{\sin^2(\phi_\ast/2f)}, \qquad
    \sigma^{(3)} = -\frac{M_{\rm Pl}^6}{f^6}\frac{\cos^2(\phi_\ast/2f)[1 - 2\sin^2(\phi_\ast/2f)]}{2\sin^2(\phi_\ast/2f)}.
\end{aligned}
\end{equation}
The inflaton field value at the end of inflation $\phi_{\rm end}$ is given by
\begin{align}
    \label{eq:phi_e_NS}
    \cos^2\frac{\phi_{\rm end}}{2f} = \frac{2f^2/M_{\rm Pl}^2}{1 + 2f^2/M_{\rm Pl}^2} \,,
\end{align}
and the relation between the number of $e$-folds $N_\ast$ and the inflaton field $\phi_\ast$ when the reference scale exited the horizon is 
\begin{align}
    \label{eq:Nast_NS}
    N_\ast = \frac{f^2}{M_{\rm Pl}^2}\ln\left|\frac{\cos^2(\phi_{\rm end}/2f)}{\cos^2(\phi_\ast/2f)}\right| \,,
\end{align}
with which the slow-roll parameters for the inflaton can be calculated. By specifying $Q_\chi$ (or $Q_\phi$), we can evaluate the spectral index, its high-order runnings, and the scalar-to-tensor ratio for this model.

%%%%%%%%%%%%%%%%%%%%%%%%%%%%%%
\subsubsection{Inverse Monomial-Spectator model}
As another example of mixed inflaton-spectator models, we consider the case with the inflaton potential given by an inverse monomial form \cite{Ratra:1987rm, Peebles:1987ek}: 
\begin{align}
    \label{eq:V_IMS}
    V(\phi) = V_0\left(\frac{\phi}{M_{\rm Pl}}\right)^{-p},\qquad p>0,
\end{align}
where $V_0$ represents the energy scale of this model. The corresponding slow-roll parameters are given by
\begin{equation}
\label{eq:SR_IMS}
\begin{aligned}
    & \epsilon = \frac{p^2}{2}\left(\frac{\phi_\ast}{M_{\rm Pl}}\right)^{-2}\,, 
     &\eta& = p(p+1)\left(\frac{\phi_\ast}{M_{\rm Pl}}\right)^{-2} \,, \\[8pt]
    &\xi^{(2)} = p^2(p+1)(p+2)\left(\frac{\phi_\ast}{M_{\rm Pl}}\right)^{-4} \,,  
     & \sigma^{(3)}& = p^3(p+1)(p+2)(p+3)\left(\frac{\phi_\ast}{M_{\rm Pl}}\right)^{-6} \,.
\end{aligned}
\end{equation}
The relation between the $e$-folding number and the inflaton value at the reference scale is given by
\begin{align}
    \label{eq:e_fold_IMS}
    N_\ast = \frac{1}{2p}\left[\left(\frac{\phi_{\rm end}}{M_{\rm Pl}}\right)^2 - \left(\frac{\phi_{\ast}}{M_{\rm Pl}}\right)^2\right] \,.
\end{align}
Since inflation does not end through a violation of the slow-roll conditions in this model, an additional mechanism is required to terminate inflation. Consequently, $\phi_{\rm end}$ is not determined by the condition $\epsilon=1$ but depends on the mechanism responsible for ending inflation. For generality, we do not specify this mechanism and instead leave $\phi_{\rm end}$ as a free parameter of the model.

%%%%%%%%%%%%%%%%%%%%%%%%%%%%%%
\subsubsection{$\chi$-extended $R^2$ model}

As a model somewhat different from the conventional ones, we consider the $\chi$-extended $R^2$ model~\cite{Wang:2025dbj}, which is a multi-field extension of the $R^2$ model. In this model, inflation consists of two stages. In the first stage, $R^2$-type inflation proceeds, and following a transition, inflation along another scalar field $\chi$ occurs in the second stage. In addition, a scalar field $\chi$ with non-minimal coupling imparts a blue-tilted component to the primordial curvature fluctuation, amplifying the power spectrum at small scales. 
This amplification not only leads to the formation of primordial black holes (PBHs) but also affects $n_s$ and $\alpha_s$ on the CMB scale.

For modes $k \le k_1$ that exit the horizon before the transition from the first stage to the second one,
the primordial power spectrum can be approximated by
\begin{align}
    \label{eq:PPS_chi_R2}
    \mathcal{P}_s(k) = \mathcal{P}_s^{R^2}(k) + \mathcal{P}_s^{\rm peak}\left(\frac{k}{k_1}\right)^a,
\end{align}
where $\mathcal{P}_s^{R^2}$ is the contribution from the $R^2$ inflation, $\mathcal{P}_s^{\rm peak}$ is the peak amplitude at the small scale, and $k_1$ is the characteristic scale corresponding to the end of the first stage. $a$ represents the strength of the blue-tilt component originating from $\chi$ and has the following relation with the non-minimal coupling $\xi$:
\begin{align}
    a = {\rm Re}\left[3 - 3\sqrt{1 - \frac{16}{3}\xi}\right].
\end{align}
From Eq.~\eqref{eq:PPS_chi_R2}, we can obtain $n_s, r$ and $\alpha_s$ at the CMB pivot scale as \cite{Wang:2025dbj}
\begin{equation}
\label{eq:ns_alpha_chi_R2}
    \begin{aligned}
    n_s &\simeq 1 - \frac{2}{N_1} + \frac{\mathcal{P}_s^{\rm peak}}{\mathcal{P}_s^{\rm CMB}} e^{-aN_1} \left(\frac{2}{N_1} + a\right)\,, 
    \qquad 
    r \simeq \frac{1}{N_1^2} \,, \\
    \alpha_s &\simeq -\frac{2}{N_1^2} + \frac{\mathcal{P}_s^{\rm peak}}{\mathcal{P}_s^{\rm CMB}} e^{-aN_1} \left(\frac{2}{N_1^2} + \frac{2a}{N_1} + a^2\right),
\end{aligned}
\end{equation}
where $N_1$ is the $e$-folding number from the time the CMB pivot scale exits the horizon until the end of the first stage of inflation. Under the sharp peak approximation, $N_1$ and the PBH mass $M_{\rm PBH}$, are related by
\begin{align}
    M_{\rm PBH} \sim 10^{18}\,{\rm g}\,\exp\left[-2(N_1 - 35)\right].
\end{align}

%%%%%%%%%%%%%%%%%%%%%%%
\subsection{Predictions for the inflationary parameters}
%%%%%%%%%%%%%%%%%%%%%%%

\begin{figure}
    \centering
    \includegraphics[width=1.0\linewidth]{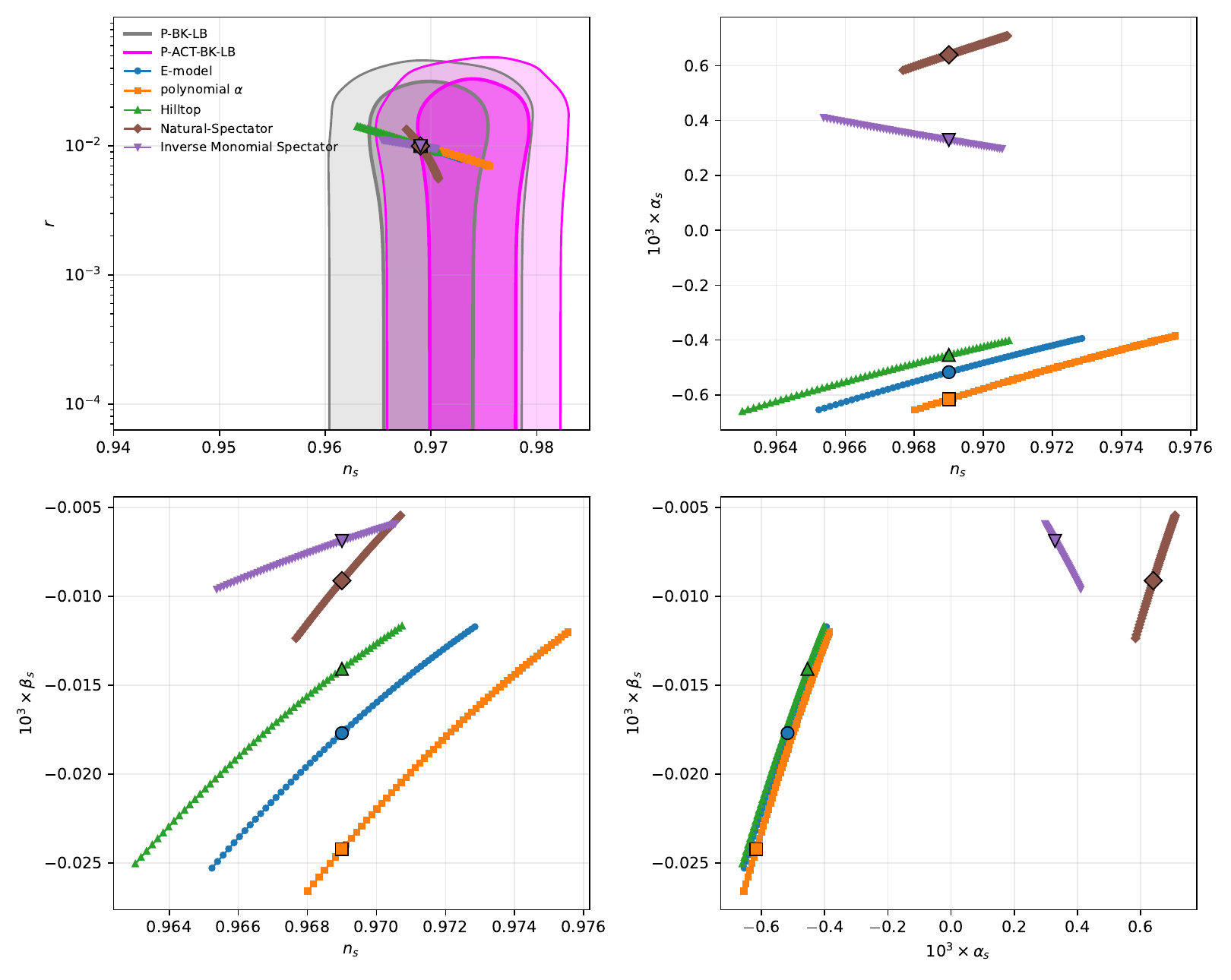}
    \caption{Plots of $n_s - r,~n_s - \alpha_s,~n_s - \beta_s,~\alpha_s - \beta_s$ predictions for the $\alpha$-attractor E-model, Polynomial $\alpha$-attractor, Hilltop, Natural-Spectator, and Inverse Monomial-Spectator models assuming some model parameters. The gray and magenta shaded regions show the $65\%$ and $95\%$ CL contours for the P-BK-LB and P-ACT-LB-BK datasets, respectively. Here, P, BK, LB, and ACT denote Planck \cite{Planck:2019nip,Pagano:2019tci}, BICEP/Keck 2018 \cite{BICEP:2021xfz}, 
    CMB lensing \cite{Carron:2022eyg,ACT:2023dou,ACT:2023kun} + BAO (DESI Year-1) \cite{DESI:2024uvr,DESI:2024lzq,DESI:2024mwx}, and the Atacama Cosmology Telescope Data Release 6 \cite{AtacamaCosmologyTelescope:2025nti}, respectively. 
    Thus, P-BK-LB corresponds to Planck + BICEP/Keck 2018 + CMB lensing + DESI Year-1 BAO, while P-ACT-LB-BK corresponds to Planck + ACT DR6 + CMB lensing + DESI Year-1 BAO + BICEP/Keck 2018. The large markers indicate cases where the spectral index $n_s$ and the scalar-to-tensor ratio $r$ take the same values across models. We adopt $50\le N_\ast \le 65$.}
    \label{fig:ns_r_alphas_betas_some_models}
\end{figure}

In this subsection, we present the predictions for $n_s$, $\alpha_s$, $\beta_s$, and $r$ for the inflation models introduced in the previous subsection, shown in the $n_s$–$r$, $n_s$–$\alpha_s$, $n_s$–$\beta_s$, and $\alpha_s$–$\beta_s$ planes, which are presented in Fig.~\ref{fig:ns_r_alphas_betas_some_models}. We set the model parameters and $N_\ast$ such that they are consistent with observational constraints and yield the same values for the spectral index $n_s$ and the scalar-to-tensor ratio $r$ across the models: $\alpha$-attractor E-model ($\alpha=4.49, m=0.50$), Polynomial $\alpha$-attractor ($\mu=2.39$), Hilltop model ($p=6.92, \mu=22.44$), Natural-spectator model ($f=4.21, Q_\phi = 0.48$), and Inverse Monomial-Spectator model ($p = 6.00, Q_\phi = 0.04$). $N_\ast$ is varied in the range of $50 \le N_\ast \le 65$.

In the top left panel of Fig.~\ref{fig:ns_r_alphas_betas_some_models}, we show the model predictions for the chosen model parameters and range of $N_\ast$, together with the current observational constraints in the $n_s-r$ plane. We consider two sets of observational constraints: those from the combination of Planck \cite{Planck:2019nip,Pagano:2019tci} + BICEP/Keck 2018 \cite{BICEP:2021xfz} + CMB lensing \cite{Carron:2022eyg,ACT:2023dou,ACT:2023kun} + DESI Y1 BAO \cite{DESI:2024uvr,DESI:2024lzq,DESI:2024mwx} (P-BK-LB), and those obtained by additionally including the Atacama Cosmology Telescope Data Release 6 (ACT DR6) \cite{AtacamaCosmologyTelescope:2025nti} (P-ACT-BK-LB). While we set model parameters such that these models yield nearly identical values in the $n_s$-$r$ plane, differences between the models emerge in the predictions for the running parameters $\alpha_s$ and $\beta_s$, which can be clearly seen in the $n_s$-$\alpha_s$, $n_s$-$\beta_s$, and $\alpha_s$-$\beta_s$ planes. Notably, the Natural-Spectator and Inverse Monomial-Spectator models exhibit $\alpha_s > 0$, while other three models predict $\alpha_s<0$. 

In the next section, we discuss whether it is possible to distinguish between inflation models that are indistinguishable in the $n_s$-$r$ plane, as shown in Fig.~\ref{fig:ns_r_alphas_betas_some_models}, from future constraints on the running parameters by utilizing the 21\,cm line power spectrum from the Dark Ages.

%%%%%%%%%%%%%%%%%%%%%%%%%%%%%%%%%%%%%%%%%
\section{Expected constraints on the runnings from dark age 21\,cm power spectrum \label{sec:results}}
%%%%%%%%%%%%%%%%%%%%%%%%%%%%%%%%%%%%%%%%%

Now in this section, we investigate expected constraints on the inflationary parameters such as the spectral index $n_s$, the running, and the running of the running parameters $\alpha_s$ and $\beta_s$, and argue that the dark age 21\,cm power spectrum will be a very powerful probe of the inflationary cosmology. 

%%%%%%%%%%%%%%%%%%%%%%%%%%%%%%%%%%%%%%%%%
\subsection{Fisher Analysis \label{sec:Fisher}}
%%%%%%%%%%%%%%%%%%%%%%%%%%%%%%%%%%%%%%%%%

To evaluate expected constraints from future 21\,cm observations on the parameters characterizing the scale-dependence of the primordial curvature power spectrum, namely $n_s$, $\alpha_s$, and $\beta_s$, we perform a Fisher matrix analysis. We also consider a combination of CMB data and future 21\,cm observations. Therefore, we include the Planck CMB specifications in our Fisher matrix analysis, even though actual Planck data are already available. As we will show, the parameter uncertainties obtained from the Fisher matrix analysis using the Planck specifications are in good agreement with those derived from the actual Planck data \cite{Planck:2018vyg}.

In this subsection, we describe the method of the Fisher matrix analysis for the power spectrum of 21\,cm  brightness temperature fluctuations from the Dark Ages \cite{Mao:2008ug,Chen:2016zuu,Cole:2019zhu,deKruijf:2024voc} and CMB \cite{Tegmark:1996bz,Zaldarriaga:1997ch,Kohri:2013mxa}.

%%%%%%%%%%%%%%%%%%%%%%
\subsubsection{21\,cm line Fisher Analysis \label{sec:21cm_fisher}}

First we describe the Fisher matrix analysis for the 21\,cm power spectrum. Since the redshifted 21\,cm signal depends not only wavenumber $k$, but also redshift $z$, 
we construct the total Fisher matrix by summing over both redshift and wavenumber bins. The Fisher matrix is written as
\begin{align}
    \label{eq::Fisher_matrix}
    F_{ij}^{21{\rm cm}}=\sum_{z}\sum_{k}\frac{1}{\delta P^2(k,z)}\frac{\partial\Delta_{21}^2(k,z)}{\partial p_i}\frac{\partial\Delta_{21}^2(k,z)}{\partial p_j},
\end{align}
where $\delta P$ is the total noise power spectrum, which is given by the sum of the contribution from the cosmic variance $\delta P_{\rm CV}$ given in Eq.~\eqref{eq:CV_error} and the thermal noise $\delta P_{\rm thermal}$ given in Eq.~\eqref{eq:thermal_noise}. 
$p_i$ denotes the cosmological parameters, specifically $p_i=(\ln A_s, n_s,\alpha_s, \beta_s)$. Other cosmological parameters, such as $\Omega_b h^2$, $\Omega_\Lambda$, and $h$, also affect the 21\,cm power spectrum. However, their effects are relatively weak, and these parameters are already tightly constrained by Planck \cite{Planck:2018vyg}. We therefore fix the cosmological parameters other than the inflationary ones to their fiducial values when calculating the Fisher matrix for the 21\,cm power spectrum, while allowing them to vary in the CMB Fisher matrix.

As already mentioned in Sec.~\ref{sec:21cmPS}, we assume the two configurations given in Table~\ref{tab:observation_spec}. 
We derive constraints on $n_s$, $\alpha_s$, and $\beta_s$ for each configuration by performing a Fisher analysis using 8 frequency bins within the frequency range $10\le\,\nu\,{\rm [MHz]}\,\le\,45$ and 11 $k$-bins within the wavenumber range $6 \times 10^{-4} \le k \, [{\rm Mpc}^{-1}] \le 700$.

For the combined analysis of CMB and  21\,cm power spectrum, we first marginalize the CMB Fisher matrix over the cosmological parameters other than the parameters of interest, $\ln A_s, n_s, \alpha_s$, and $\beta_s$. We then add this marginalized CMB Fisher matrix to the one obtained from the 21\,cm power spectrum, defined in the same parameter space, to evaluate the combined CMB+21\,cm constraints for the inflationary parameters.

%%%%%%%%%%%%%%%%%%%%%%%%%%%%%
\subsubsection{CMB Fisher Analysis \label{sec:CMB_fisher}}
Here we briefly describe the Fisher analysis for CMB, following the notation of  Ref.~\cite{Kohri:2013mxa}. The Fisher matrix for the CMB is calculated by
\begin{align}
    \label{eq:fiher_CMB}
    F_{ij}^{\rm CMB}=\sum_{\ell=\ell_{\rm min}}^{\ell_{\rm max}}\frac{2\ell+1}{2}f_{\rm sky}{\rm Tr}\left[{\bm C_{\ell}}^{-1}\frac{\partial{\bm C}_{\ell}}{\partial p_i}{\bm C}_{\ell}^{-1}\frac{\partial\bm{C}_{\ell}}{\partial p_j}\right],
\end{align}
where $\bm{C}_{\ell}$ is a covariance matrix of CMB, $f_{\rm sky}$ is the fraction of the sky effectively covered by observations, and $p_i$ represents cosmological parameters. For CMB analysis, we take $p_i =(\Omega_bh^2,~\Omega_ch^2,~H_0,~\tau,~\ln A_s,~n_s,~\alpha_s,~\beta_s)$, where $\tau$ is the optical depth (for other parameters, see the paragraph below Eq.~\eqref{eq:dimensionless_PS}). When deriving constraints on $n_s$, $\alpha_s$, and $\beta_s$, we marginalize over the other parameters.

The covariance matrix is a $2\times2$ matrix for each multipole $\ell$, involving the temperature $T$ and E-mode polarization $E$, and can be written explicitly as
\begin{align}
    \label{eq:covariance_CMB}
    \bm{C}_{\ell} =
    \begin{pmatrix}
       C_{\ell}^{TT}+N_{\ell}^{T} & C_{\ell}^{TE}  \\
       C_{\ell}^{TE} & C_{\ell}^{EE}+N_{\ell}^{P} \\
    \end{pmatrix},
\end{align}
where $C_{\ell}^{X}$ with $X=TT, EE, TE$ denotes lensed angular power spectrum. $N_{\ell}^{T}$ and $N_{\ell}^{P}$ are the temperature and polarization noise power spectra, respectively. For a single frequency, $N_{\ell}^T,~N_{\ell}^{P}$ at each frequency channel is given by
\begin{align}
    N_{\ell}^{Y}(\nu)=\Delta_Y^2\exp\left[\ell(\ell+1)\sigma_b^2(\nu)\right],~~Y=T,~P,
\end{align}
where $\sigma_b(\nu)=\theta_{\rm FWHM}/\sqrt{8\ln 2}$ denotes the width of the beam. For $\Delta_Y$ and $\theta_{\rm FWHM}$, we adopt the Planck specifications \cite{Planck:2006aa} summarized in Table~\ref{tab:Planck_specifications}. We use all three frequencies by identifying the noise power spectrum for the frequency-combined analysis as 
\begin{align}
    \left(N_{\ell}^Y\right)^{-1} = \sum_{\nu_i}\frac{1}{N_{\ell}^Y(\nu_i)}.
\end{align}
In the analysis, we assume $f_{\rm sky}=0.65$, $\ell_{\rm min}=2$, and $\ell_{\rm max}=3000$. 
As mentioned at the beginning of this section, although the actual CMB data are already available, we include the CMB information in the Fisher matrix to investigate how combining the CMB with the 21\,cm power spectrum constrains the inflationary parameters. We have verified that the CMB specifications listed in Table~\ref{tab:Planck_specifications} yield uncertainties in the cosmological parameters that are very similar to those obtained from the actual CMB data.

\begin{table}
    \centering
    \begin{tabular}{|c|c|c|c|}
         \hline
         Frequency [GHz] & $\theta_{\rm FWHM}$ [arcmin] & $\Delta_T$ [$\mu$K arcmin] & $\Delta_P$ [$\mu$K arcmin]\\ \hline\hline
         100 & 9.5 & 64.6 & 104\\ \hline
         143 & 7.1 & 42.6 & 80.9\\ \hline
         217 & 5 & 65.5 & 134\\ \hline
    \end{tabular}
    \caption{Specifications for CMB adopted in the analysis, which correspond to those for Planck \cite{Planck:2006aa}. $f_{\rm sky}=0.65$ and $\ell_{\rm max} = 3000$ are assumed.}
    \label{tab:Planck_specifications}
\end{table}

\subsection{Expected constraints from each configuration \label{sec:future_const}}

To investigate expected constraints from future observations of 21\,cm power spectrum during the Dark Ages, we have calculated the Fisher matrix for the 21\,cm power spectrum and CMB as described in the previous section. In the analysis, we assume the following fiducial values for the cosmological parameters: 
$\Omega_{b}h^2=0.02237$, $\Omega_{c}h^2=0.12$, $H_0=67.36$, $A_s=2.1\times10^{-9}, \tau=0.0544$,  which are taken from the Planck result \cite{Planck:2018vyg}.
The Helium-4 abundance $Y_p$, which is not varied in the Fisher analysis, is taken to be 
$Y_{p}=0.2436$ \cite{Hsyu:2020uqb}. The fiducial values for $n_s$, $\alpha_s$, and $\beta_s$ are assumed as $n_s = 0.9690$, $\alpha_s = -4.5\times10^{-4}$, and $\beta_s = -1.4\times10^{-5}$, respectively, corresponding to the Hilltop model with $N_\ast = 60.93$ shown in Fig.~\ref{fig:ns_r_alphas_betas_some_models}.
The reference scale for the primordial power spectrum is taken to be $k_{\ast}= k_{\rm planck} = 0.05~{\rm Mpc}^{-1}$.
As mentioned in the previous section, when combining the 21\,cm power spectrum with the CMB, we first calculate the CMB Fisher matrix and marginalize over all cosmological parameters except the inflationary ones, $\ln A_s$, $n_s$, $\alpha_s$, and $\beta_s$. We then add the Fisher matrix from the 21\,cm power spectrum to obtain the expected constraints from future 21\,cm observations.

\begin{figure}
    \centering
    \includegraphics[width=1.0\linewidth]{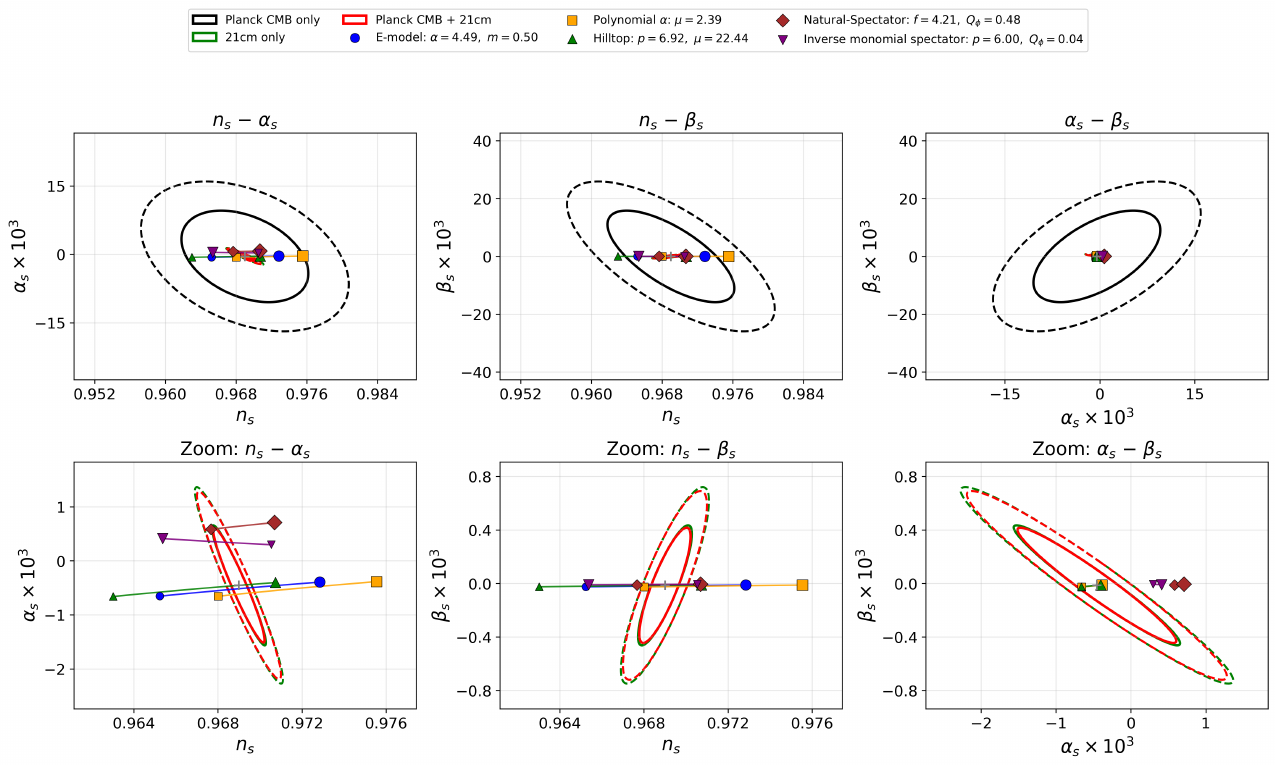}
    \caption{Plots of $n_s-\alpha_s$, $n_s - \beta_s$, and $\alpha_s - \beta_s$ constraints in the configuration A. The markers and the solid lines connecting them represent predictions for $n_s$, $\alpha_s$, and $\beta_s$ based on the five inflationary models discussed in Sec.~\ref{sec:inflation}. In these models, the model parameters have been adjusted so that the values of $n_s$ and $r$ match, as introduced in Sec.~\ref{subsec:example_model}, and $N_\ast$ is varied over the range $50 \le N_\ast \le 65$.}
    \label{fig:ns_alphas_betas_configA}
\end{figure}

\begin{figure}
    \centering
    \includegraphics[width=1\linewidth]{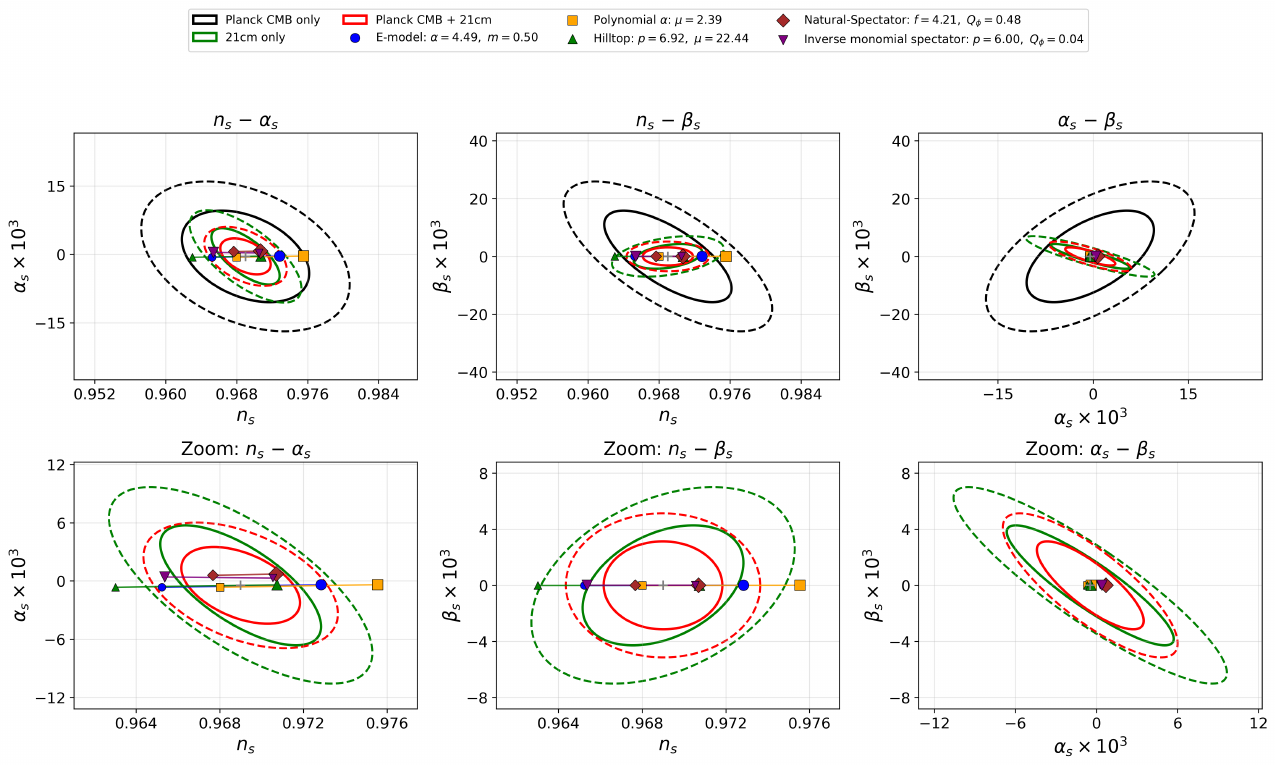}
    \caption{Plots of $n_s-\alpha_s$, $n_s - \beta_s$, and $\alpha_s - \beta_s$ constraints in the configuration B. The markers and the solid lines connecting them represent the predicted values of $n_s$, $\alpha_s$, and $\beta_s$ based on the five inflationary models. In these models, the model parameters have been adjusted so that the values of $n_s$ and $r$ match, as introduced in Sec.~\ref{subsec:example_model}, and $N_\ast$ is varied over the range $50 \le N_\ast \le 65$.}
    \label{fig:ns_alphas_betas_configB}
\end{figure}

\begin{table}
    \centering
    \begin{tabular}{|c|c|c||c|} \hline
         & configuration A & configuration B  & Planck\\ \hline
         $1\sigma(n_s)$ & $8.4\times10^{-4}$ & $2.5\times10^{-3}$ & $4.3\times10^{-3}$ \\ \hline
         $1\sigma(\alpha_s)$ & $7.3\times10^{-4}$ & $4.1\times10^{-3}$ & $9.9\times10^{-3}$ \\ \hline
         $1\sigma(\beta_s)$ & $3.0\times10^{-4}$ & $2.8\times10^{-3}$ & $1.2\times10^{-2}$ \\ \hline
    \end{tabular}
    \caption{$1\sigma$ uncertainties in $n_s$, $\alpha_s$, and $\beta_s$ from the analysis assuming the 21\,cm power spectrum alone (configurations A and B) or the CMB (Planck) alone. }
    \label{tab:1sigma_eachconfig}
\end{table}

\begin{table}
    \centering
    \begin{tabular}{|c|c|c|c|} \hline
         & $1\sigma(n_s)$ & $1\sigma(\alpha_s)$ & $1\sigma(\beta_s)$ \\ \hline
         configuration A + Planck & $8.0\times10^{-4}$ & $7.0\times10^{-4}$ & $2.8\times10^{-4}$ \\ \hline
         configuration B + Planck  & $1.9\times10^{-3}$ & $2.6\times10^{-3}$ & $2.1\times10^{-3}$\\ \hline
    \end{tabular}
    \caption{$1\sigma$ uncertainties in $n_s$, $\alpha_s$, and $\beta_s$ from the analysis in combination with CMB (Planck)  and the 21\,cm power spectrum for each observational configuration. }
    \label{tab:1sigma_Planck_21cm}
\end{table}

\begin{figure}
    \centering
    \includegraphics[width=1\linewidth]{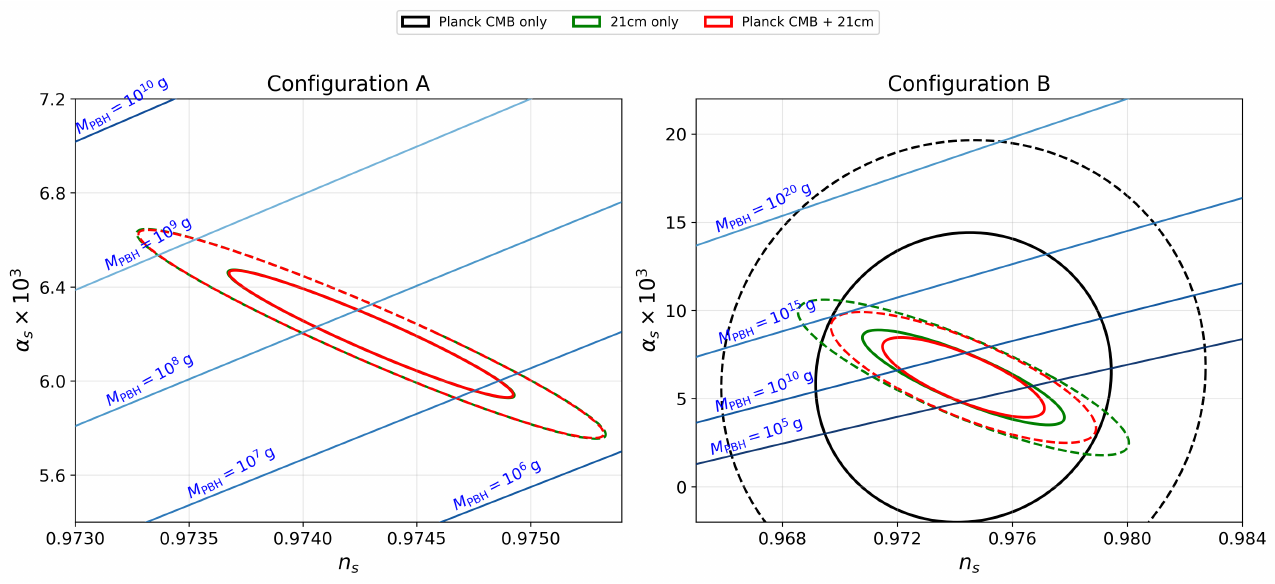}
    \caption{Predictions of the $\chi$-extended $R^2$ inflation model in the $n_s$--$\alpha_s$ plane for various values of $M_{\rm PBH}$, obtained from Eq.~\eqref{eq:ns_alpha_chi_R2}. The left and right panels show the expected constraints from 21\,cm power-spectrum observations for Configurations A and B, respectively. In each panel, we also show the constraints obtained by combining the 21\,cm observations with CMB (Planck) data.
}
    \label{fig:ns_alphas_PBH_configA_B}
\end{figure}

Figures~\ref{fig:ns_alphas_betas_configA} and \ref{fig:ns_alphas_betas_configB} show the constraints on $n_s$, $\alpha_s$, and $\beta_s$ from the Dark Ages 21\,cm power spectrum for configurations A and B, respectively. 
For each configuration, we show the constraints from the 21\,cm power spectrum alone, from the CMB alone, and from their combination. 
We also plot the predictions of five inflationary models in each parameter plane: the $\alpha$-attractor E-model, Polynomial $\alpha$-attractor, Hilltop, Natural-Spectator, and Inverse Monomial-Spectator models.
For configuration A, the constraints from the 21\,cm power spectrum alone are sufficiently tight to dominate the combined constraints.
As shown in Fig.~\ref{fig:ns_alphas_betas_configA}, the adopted inflationary models are largely degenerate in their predictions for $n_s$ and $r$, but can generally be distinguished by precise measurements of $n_s$, $\alpha_s$, and $\beta_s$ from the Dark Ages 21\,cm power spectrum. The main exception is the case where the predictions for $\alpha_s$ and $\beta_s$ remain degenerate. In particular, the Natural-Spectator and Inverse Monomial-Spectator models can be clearly distinguished from the other three models because they predict a positive $\alpha_s$, whereas the other three models predict a negative $\alpha_s$.

On the other hand, for configuration B, although future 21\,cm observations can significantly improve the constraints on $n_s$, $\alpha_s$, and $\beta_s$ compared with those from Planck, they may not be sufficient to distinguish among the 5 inflationary models shown in Figs.~\ref{fig:ns_alphas_betas_configA} and \ref{fig:ns_alphas_betas_configB}. Nevertheless, it would be still helpful to test some particular models such as the $\chi$-extended $R^2$ inflation model discussed in Sec.~\ref{sec:inflation}. 
Figure~\ref{fig:ns_alphas_PBH_configA_B} illustrates the predictions for $n_s$ and $\alpha_s$ in a model with $\mathcal{P}_s^{\rm peak}/\mathcal{P}_s^{\rm CMB}=10^7$, which gives a sizable present-day PBH abundance of $f_{\rm PBH}\sim 1$. Following Ref.~\cite{Wang:2025dbj}, we consider several values of the PBH mass for illustration. For configuration A (left panel), we take $M_{\rm PBH}=10^6$, $10^7$, $10^8$, $10^9$, and $10^{10}~{\rm g}$, while for configuration B (right panel), we take $M_{\rm PBH}=10^5$, $10^{10}$, $10^{15}$, and $10^{20}~{\rm g}$. Each choice of $M_{\rm PBH}$ fixes the corresponding value of $N_1$, which in turn determines the predictions for $n_s$ and $\alpha_s$. Expected constraints from future 21\,cm power-spectrum observations, both alone and in combination with Planck data, are also shown in the figure. To compare these expected constraints with the current constraints from P-ACT-LB~\cite{AtacamaCosmologyTelescope:2025nti}, we perform a Fisher analysis for Fig.~\ref{fig:ns_alphas_PBH_configA_B} that excludes $\beta_s$. We adopt fiducial values of $n_s = 0.9743$ and $\alpha_s = 0.0062$, which differ from those used in the other analyses in this paper.

Current CMB observations already place constraints on $M_{\rm PBH}$ through the constraints on $n_s$ and $\alpha_s$. Future 21\,cm observations can provide a much more stringent test, even for Configuration~B, as shown in the right panel of Fig.~\ref{fig:ns_alphas_PBH_configA_B}. For Configuration~A, the constraints are significantly improved, as shown in the left panel. In particular, PBHs with $M_{\rm PBH} \lesssim 10^{6}~{\rm g}$ or $M_{\rm PBH} \gtrsim 10^{10}~{\rm g}$ may be excluded as dark matter candidates in this model at the $2\sigma$ level.

Table~\ref{tab:1sigma_eachconfig} shows the $1\sigma$ uncertainties on $n_s$, $\alpha_s$, and $\beta_s$ for each 21\,cm observation configuration. As shown in the table, Configuration~B provides constraints comparable to those from Planck, while Configuration~A provides tighter constraints. Table~\ref{tab:1sigma_Planck_21cm} shows the $1\sigma$ uncertainties on $n_s$, $\alpha_s$, and $\beta_s$ obtained by combining the Planck CMB constraints with those from the Dark Ages 21\,cm power spectrum. In particular, when combining 21\,cm power spectrum with CMB data, configuration~B can further tighten the constraints.

%%%%%%%%%%%%%%%%%%%%%%%%%%%%%%
\subsection{Dependence on the pivot scale \label{sec:pivot}}

\begin{figure}
    \centering
    \includegraphics[width=1\linewidth]{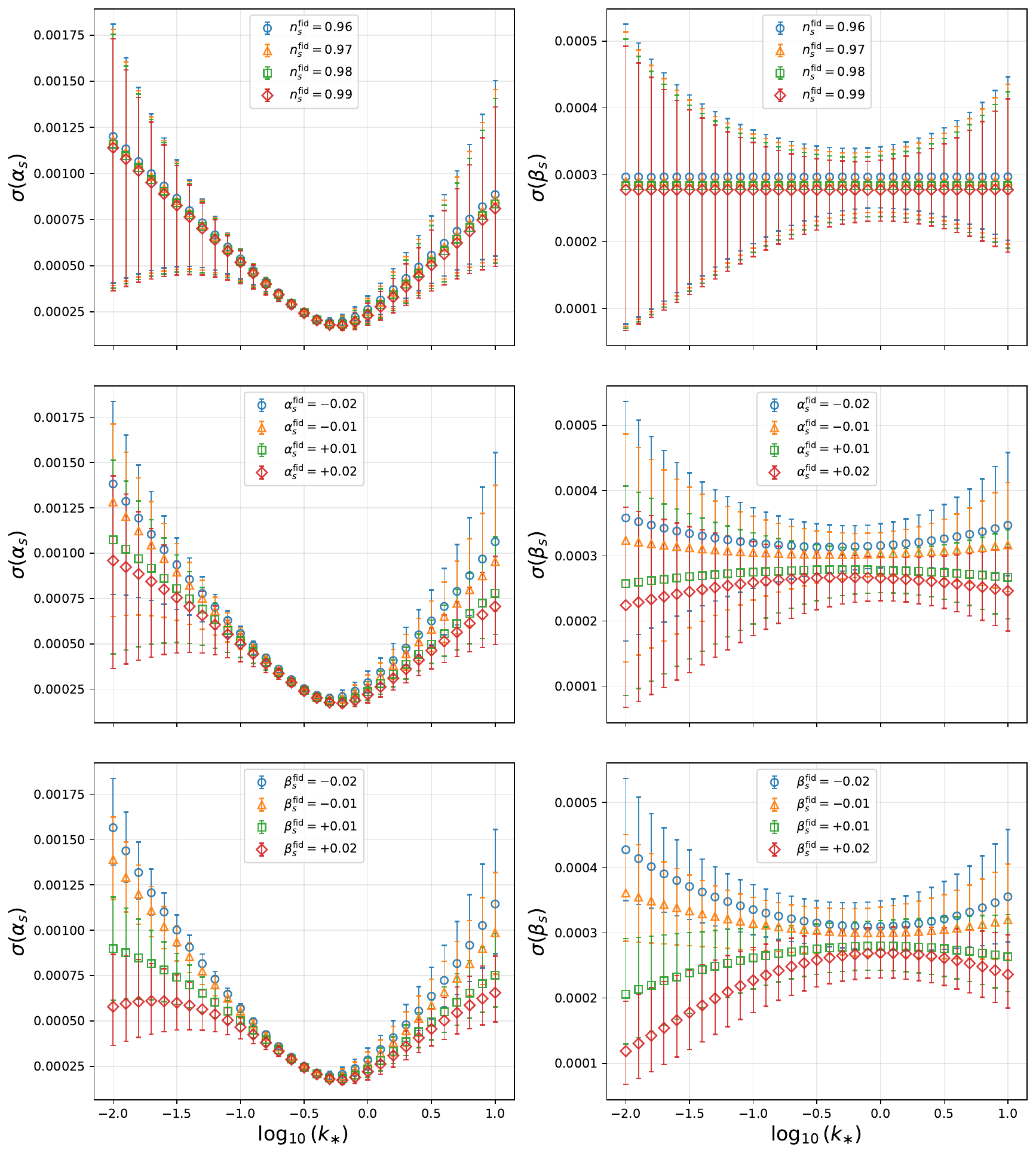}
    \caption{Plots of $1\sigma$ uncertainties on $\alpha_s$ (left panels) and $\beta_s$ (right panels) as functions of the pivot scale $k_\ast$ for observational Configuration A. The top, middle, and bottom panels show results for different fiducial values of $n_s$, $\alpha_s$, and $\beta_s$, respectively. Different colors denote the fiducial values indicated in each panel. For each $k_\ast$, the points represent the median uncertainty obtained by varying the remaining fiducial parameters over the ranges considered, and the error bars span the corresponding minimum and maximum values.}
    \label{fig:fisher_kpiv_sigma_alphas_betas_configA}
\end{figure}

\begin{figure}
    \centering
    \includegraphics[width=1\linewidth]{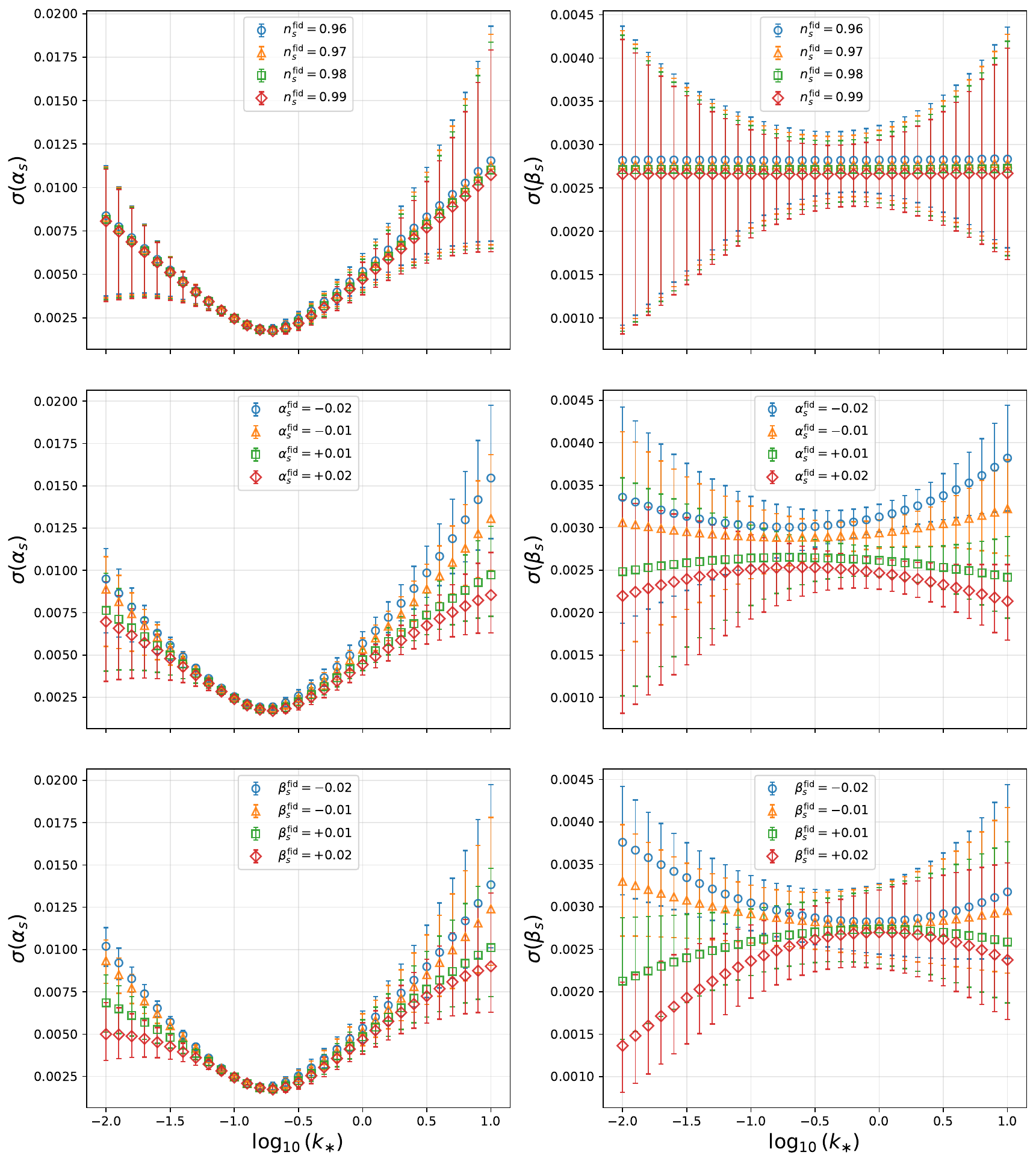}
    \caption{Same as Fig.~\ref{fig:fisher_kpiv_sigma_alphas_betas_configA}, but for Configuration~B.
    }
    \label{fig:fisher_kpiv_sigma_alphas_betas_configB}
\end{figure}

In the previous section~\ref{sec:future_const}, we adopted the reference scale $k_{\ast}=k_{\rm Planck}=0.05~{\rm Mpc}^{-1}$, corresponding to the CMB scale. However, as shown in Fig.~\ref{fig:PS_noise}, the Dark Ages 21\,cm power spectrum is sensitive to smaller scales than the CMB. In particular, the observational configurations considered in this paper can probe scales with $k \gtrsim O(10^{-1})~{\rm Mpc}^{-1}$. Therefore, the reference scale $k_{\ast}=0.05~{\rm Mpc}^{-1}$ may not be optimal for analyzing the expected constraints with the 21\,cm power spectrum. In this section, we examine how the constraints on the running and running of the running of the spectral index, $\alpha_s$ and $\beta_s$, depend on the choice of the pivot scale $k_{\ast}$. We also investigate how the fiducial values of $A_s$, $n_s$, $\alpha_s$, and $\beta_s$ affect the resulting constraints.

Figures~\ref{fig:fisher_kpiv_sigma_alphas_betas_configA} and \ref{fig:fisher_kpiv_sigma_alphas_betas_configB} show the $1\sigma$ uncertainties in $\alpha_s$ (left panels) and $\beta_s$ (right panels) as a function of $k_\ast$ for Configurations~A and B, respectively, obtained by varying $n_s$, $\alpha_s$, and $\beta_s$. Here the analysis only includes the 21\,cm power spectrum. Note that, when the fiducial values of $n_s, \alpha_s$ and $\beta_s$ are varied, the one for $A_s$ is changed accordingly with the following formula:
\begin{align}
\label{eq:As_fid}
    A_s^{\rm fid}=A_s^{(k_{\rm Planck})}\exp{\left[(n_{s}^{{\rm fid}}-1)\ln\left(\frac{k_{\ast}}{k_{\rm Planck}}\right) + \frac{1}{2}\alpha_{s}^{{\rm fid}}\ln^2\left(\frac{k_{\ast}}{k_{\rm Planck}}\right)+\frac{1}{6}\beta_{s}^{{\rm fid}}\ln^3\left(\frac{k_{\ast}}{k_{\rm Planck}}\right)\right]},
\end{align} 
where $A_s^{(k_{\rm Planck})} = 2.1 \times 10^{-9}$ is the fiducial value for $A_s$ when $k_\ast = k_{\rm Planck}$.  $n_{s}^{{\rm fid}},\alpha_{s}^{{\rm fid}}$, and $\beta_{s}^{{\rm fid}}$ are their fiducial values. 
In each figure, the value of the fixed variable is indicated in the legend of the corresponding panel. The variables not shown in the legend are varied within the ranges $0.95 \leq n_s \leq 0.99$, $-0.02 \leq \alpha_s \leq 0.02$, and $-0.02 \leq \beta_s \leq 0.02$. This variation results in an error bar for each value of $k_\ast$ shown in the figures.
For example, in the top-left panel of Fig.~\ref{fig:fisher_kpiv_sigma_alphas_betas_configA}, $n_s$ is fixed to the values indicated in the legend, while $\alpha_s$ and $\beta_s$ are varied over the ranges given above. The fiducial value of $A_s$ is determined automatically according to Eq.~\eqref{eq:As_fid}. The central value for each $n_s^{\rm fid}$ and $k_\ast$ is indicated by a marker (circle, triangle, square, or diamond).
The error bar at each point represents the uncertainty in $\sigma(\alpha_s)$ arising from the variation of the fiducial values of $\alpha_s$ and $\beta_s$.

From Figs.~\ref{fig:fisher_kpiv_sigma_alphas_betas_configA} and \ref{fig:fisher_kpiv_sigma_alphas_betas_configB}, we find that $1\sigma$ uncertainties for $\alpha_s,~\beta_s$ change by a factor of a few depending on the reference scale $k_{\ast}$. Also, $1\sigma$ errors depend on the fiducial values of $n_s, \alpha_s$ and $\beta_s$ and become more pronounced at smaller or larger $k_{\ast}$. For Configuration~A, shown in Fig.~\ref{fig:fisher_kpiv_sigma_alphas_betas_configA}, the minimum uncertainty in $\alpha_s$, $\sigma(\alpha_s)$, is obtained at $k_{\ast}\sim0.63~{\rm Mpc}^{-1}$, and this value is almost independent of the fiducial values of $n_s$, $\alpha_s$, and $\beta_s$. For Configuration~B, the reference scale corresponding to the minimum $\sigma(\alpha_s)$ shifts to $k_{\ast}\sim0.2~{\rm Mpc}^{-1}$. This difference arises because the wavenumber at which thermal noise becomes dominant depends on the observational configuration.

On the other hand, the minimum uncertainty in $\beta_s$, $\sigma(\beta_s)$, shows only a weak dependence on $k_\ast$ when the fiducial values of $n_s$ and $\alpha_s$ are varied, whereas it depends more strongly on the fiducial value of $\beta_s$. As shown in the top-right panels of Figs.~\ref{fig:fisher_kpiv_sigma_alphas_betas_configA} and \ref{fig:fisher_kpiv_sigma_alphas_betas_configB}, varying $n_s^{\rm fid}$ has little effect on $\sigma(\beta_s)$: although the uncertainty in $\sigma(\beta_s)$ changes with $k_\ast$, its central value remains almost unchanged. A similar behavior is found when $\alpha_s^{\rm fid}$ is varied, although $\sigma(\beta_s)$ shows a mild dependence on $k_\ast$ and $\alpha_s^{\rm fid}$. In contrast, varying $\beta_s^{\rm fid}$ has a much stronger effect on $\sigma(\beta_s)$. As seen in the bottom-right panels, this strong dependence explains why the error bars in the top-right and middle-right panels are relatively large, even though the central values of $\beta_s^{\rm fid}$ do not change significantly.

%%%%%%%%%%%%%%%%%%%%%%
\section{Conclusion \label{sec:conclusion}}
%%%%%%%%%%%%%%%%%%%%%%

In this paper, we have investigated how precisely future observations of the Dark Ages 21\,cm power spectrum, combined with CMB (Planck) observations, can probe the scale dependence of the primordial power spectrum. Specifically, we have derived expected constraints on the spectral index $n_s$, its running $\alpha_s$, and its running of the running $\beta_s$ using a Fisher analysis for two observational configurations, A and B, listed in Table~\ref{tab:observation_spec}. Configuration~A can provide stringent constraints on $n_s$, $\alpha_s$, and $\beta_s$, and would help distinguish between inflation models that are difficult to differentiate in the $n_s$--$r$ plane, particularly models that predict different signs of $\alpha_s$ (see Fig.~\ref{fig:ns_alphas_betas_configA}).

Although Configuration~B may not be sufficient to distinguish among conventional slow-roll inflation models, combining 21\,cm data with CMB observations can yield tighter constraints on $n_s$, $\alpha_s$, and $\beta_s$ than CMB data alone, as the combination helps break parameter degeneracies. At the same time, Configuration~B can provide stringent tests of models with unconventional predictions. As an example, we considered the $\chi$-extended $R^2$ inflation model~\cite{Wang:2025dbj}, which can accommodate PBH formation. As shown in Fig.~\ref{fig:ns_alphas_PBH_configA_B}, even Configuration~B can place strong constraints on the PBH mass. Since the Dark Ages 21\,cm power spectrum probes much smaller scales than the CMB, it can provide a powerful test of models that predict deviations from the standard primordial power spectrum on small scales.

We have also investigated the optimal pivot scale and the fiducial values of $n_s$, $\alpha_s$, and $\beta_s$ that yield small uncertainties in $\alpha_s$ and $\beta_s$ in Sec.~\ref{sec:pivot}. Figures~\ref{fig:fisher_kpiv_sigma_alphas_betas_configA} and \ref{fig:fisher_kpiv_sigma_alphas_betas_configB} show that the $1\sigma$ uncertainty in $\alpha_s$ depends on the choice of pivot scale. In contrast, the uncertainty in $\beta_s$ is more sensitive to the fiducial value of $\beta_s$. The choice of pivot scale also affects the uncertainty in $\beta_s$, although its impact is less significant than for $\alpha_s$. Importantly, the pivot scale that minimizes the uncertainty in $\alpha_s$ depends on the observational configuration. Therefore, when using the Dark Ages 21\,cm power spectrum to constrain inflationary parameters, particularly in combination with other datasets, the dependence of the parameter uncertainties on the choice of pivot scale and fiducial parameter values should be carefully taken into account.

As argued in this paper, future Dark Ages 21\,cm observations would open a new window into the primordial Universe by probing scales far beyond those accessible to the CMB. Such observations may provide a powerful means of distinguishing inflationary scenarios and testing models with nontrivial small-scale features in the primordial power spectrum.

\section*{Acknowledgements}
This work was partially supported by JSPS KAKENHI Grant Numbers JP25K01004 (TT), JP23H00108 (SY), and JP24K00627 (SY), and MEXT KAKENHI JP25H01543 (TT), JP26H00402 (TT). F.O.~was supported by individual research funding from Nihon University.

\clearpage 
\bibliography{refs}

\end{document}